\documentclass[pre, aps, twocolumn, floatfix, superscriptaddress]{revtex4}
\usepackage{graphicx,color}

\begin{document}

\title{Electrocharged facepiece respirator fabrics using common materials}
\author{M. M. Bandi}
\affiliation{Nonlinear and Non-equilibrium Physics Unit, OIST Graduate University, Onna, Okinawa, 904 0495 Japan}
\email[Email: ]{bandi@oist.jp} 

\date{\today}

\begin{abstract}{
Face masks in general, and N95 filtering facepiece respirators (FRs) that protect against SARS-Cov-2 virion in particular, have become scarce during the ongoing COVID-19 global pandemic. This work presents practical design principles for the fabrication of electrocharged filtration layers employed in N95 FRs using commonly available materials and easily replicable methods. The input polymer is polypropylene or polystyrene, and can include discarded plastic containers of these materials, and the fabrication setup is based on the cotton candy (CC) principle. The primary parameters underlying the CC principle are translated to simple design rules that allow anyone to construct their own fabrication system from common parts, or employ a commercial cotton candy machine with minimal modifications. Finally, basic characterization results for structural and filtration properties of electrocharged fabrics made using the CC principle are detailed.
}
\end{abstract}
\maketitle

\section{Introduction}
The ongoing COVID-19 pandemic has witnessed multi-fold increase in face mask use for protection against viral infection, with many countries now mandating face masks in public areas \cite{CDCFMask}. This sudden demand surge has created a scarcity of face masks, necessitating homemade cloth mask fabrication \cite{CDCFMask}. But neither homemade cloth masks nor surgical masks are designed to protect against the SARS-Cov-2 virion, only respirators conforming to N95 or higher standard are rated for such protection. This capability of N95 filtering facepiece respirators (FRs) is owed primarily to an electrocharged filtration layer among other notable design features, deemed the most efficient among various particle filtration methods employed in face masks. 

Taken together, these features permit filtration of $\ge 95\%$ of particles of size $\ge 0.3 \mu$m under test standards designated by the US National Institute for Occupational Safety and Health (NIOSH) \cite{NIOSH-N95} for N95 (US: NIOSH-42C-FR84) and its counterparts, including FFP2 (Europe: EN149-2001), KN95 (China:  GB2626-2006), P2 (Australia and New Zealand: AS/NZ1716:2012), Korea 1st class (South Korea: KMOEL-2017-64), and DS2 (Japan: JMHLW-Notification214 from year 2018) facepiece respirators \cite{3M}. Unfortunately, the electrocharged polymer filtration layers used in these FRs are manufactured through industrially sophisticated  processes that are hard to duplicate using commonly available materials or methods.

This article details a process to fabricate electrocharged polymer based fabric using commonly available materials and easily replicable methods. The fabrication setup is based on the Cotton Candy (CC) principle, also known as rotary jet spinning or centrifugal spinning method \cite{RSpinReview}. The primary control parameters for the CC method \cite{RJet2} are translated to practical design rules for either the construction of a fabric manufacturing setup from common parts, or through minimal modification of commercial cotton candy machines. Practical solutions to tune the control parameters for fabrication of electrocharged polymer fabrics are also specified. Finally, characterisation of electrocharged fabrics made using the CC method for structural as well as filtration properties using two mask designs are presented.

\section{Filtration Principles}
\subsection{Basic Mechanisms}
Face mask filtration mechanisms must optimize between two competing requirements. On one hand the mask's filtration layers must possess an average pore diameter small enough to trap and filter particles from being inhaled, but at the same time too small a pore diameter prevents the user from breathing comfortably \cite{Revoir1995}. For this reason, face mask filtration layers cannot be fabricated below a certain pore diameter. Furthermore, mechanisms for filtration of large particles differ from those for small particles, thereby requiring a range of filtration strategies to be adopted. These strategies commonly involve three physical mechanisms, {\it viz.} inertial impaction, diffusion, and electrostatic attraction \cite{NAS2006}.

Large particles with diameters $\ge 1 \mu$m possess inertia to deviate from aerodynamic streamlines and collide with filtration fibers and get caught in the filter mesh. Small particles of typical diameters $\le 0.1 \mu$m which follow streamlines while undergoing diffusion, execute a complex, meandering trajectory through the tortuous porous matrix and get trapped in the filtration layers. These two mechanisms are easily achieved in any cloth-based or commercial (surgical or PM2.5) masks, but they do not incorporate the third mechanism of electrostatic filtration \cite{Konda, Davies}, which traps particles of intermediate sizes in the range $0.1 - 1 \mu$m and is considered most effective of the three mechanisms. When an electrostatically charged layer is embedded among standard mask filtration layers, oppositely charged particles (both small and large in diameter) are attracted by the long-range electrostatic Coloumb force towards the electrocharged layer. Once caught, the particles are held in place through van der Waals forces.

\subsection{Electrocharged Filtration}
It is known from common experience, especially in cold climates with low ambient humidity, that when two dissimilar fabrics rub against each other they gain static electricity, a phenomenon known as triboelectric charging. Fabrics woven from natural fibers like wool or cotton, which possess high roughness, and even synthetic fabrics like Nylon are common examples of fabrics with high triboelectric charging ability. The idea of exploiting charged fabrics to aid in filtration goes back nearly five decades \cite{Frederick1974}, and indeed some early face mask designs incorporating electrocharged filtration employed wool or felt fibers, with resin additives to enhance filtration efficiency many times over that achieved with basic fabric materials alone \cite{NAS2006}. However, resin additives degrade upon exposure to airborne oil aerosol droplets, which can shield electrostatic charges. Consequently over time, synthetic electret fabrics such as plastic fibers (e.g. polypropylene and polystyrene) with high electrostatic charge characteristies were found to resist the shielding effect of oil aerosols quite effectively and came to be adopted as the material of choice for electrocharged filtration fabrics. An additional advantage of plastic electret fabrics was that they were non-woven, thus saving fabrication time. Finally, the disordered fabric pattern in non-woven electret fabrics, as opposed to knitted fabric with interleaving fiber strands in a grid, assures a highly irregular porous medium ensuring particles follow tortuous streamlines through the porous matrix.

\section{Electrocharged material fabrication methods}
All plastic electrets or electrocharged fabrics are manufactured by bonding polymer fibers in a porous mesh. In order to generate polymer fibers, one starts with liquid polymer either from solution (polymer dissolved in solvent) or polymer melt. The bulk liquid must be forced through a tiny orifice to overcome capillary forces, and quickly accelerated to stretch or extend the viscoelastic solution into fibers of diameter ranging from 100s of nanometers to 10s of $\mu$m. This dispersity in fabric diameters helps provide large surface area exposure (relative to bulk volume) to attract particles. Additionally, the non-woven fibers bonding together with the dispersity helps increase fabric disorder and results in a tortuous porous matrix. These fibers either evaporate their solvent or cool if derived from melt, as they traverse from the orifice to be collected on a surface where they solidify into an enmeshed fabric. The various fabrication methods therefore vary in the forcing mechanisms employed to overcome capillary force and thence accelerate the polymer. The three primary methods, {\it viz.} Electrospinning, Melt Blowing, and Rotary Jet Spinning, are briefly reviewed as they inform the design rules to follow subsequently.

\begin{figure*}
\begin{center}
\includegraphics[width= 1.75\columnwidth]{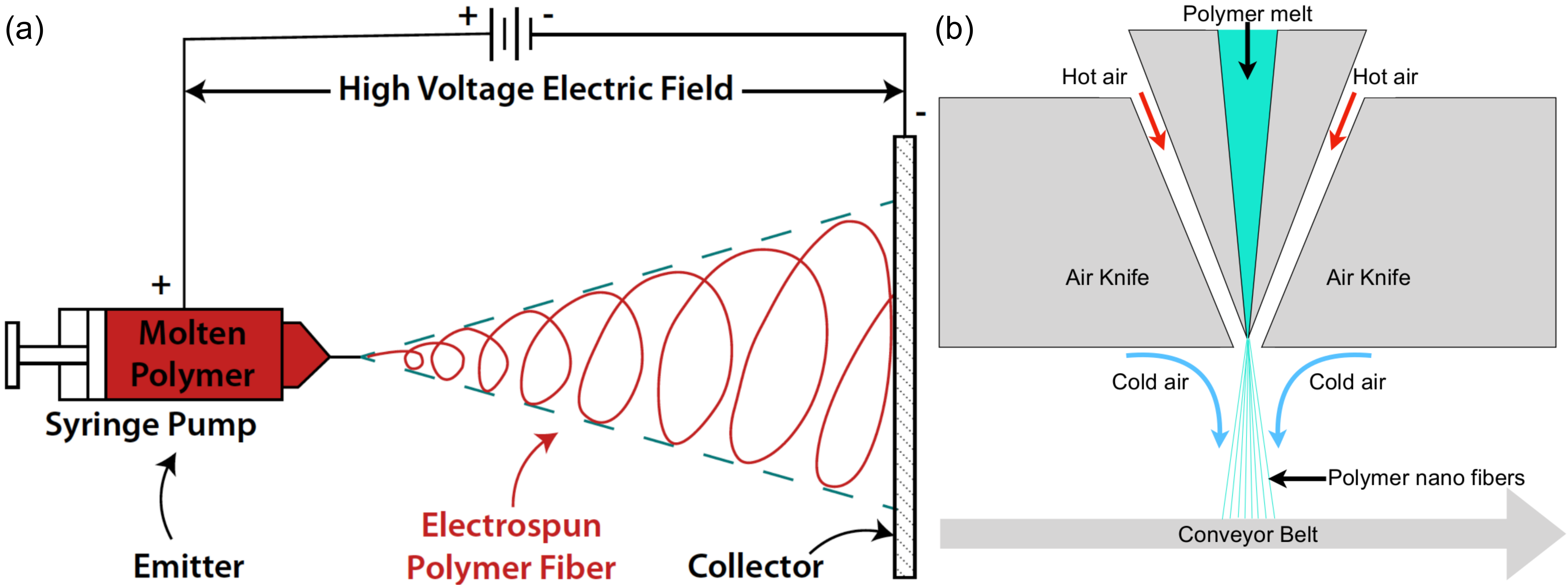}
\end{center}
\caption{(Color online) Schematic of (a) a generic electrospinning setup and (b) cross-sectional view of a Melt Blowing setup.}
\label{fig1}
\end{figure*}

{\it Electrospinning:} Electrospinning is a widely used platform for generating polymer fibers \cite{ESpinRev1, ESpinRev2, ESpinRev3, ESpinRev4, ESpinRev5}. In this method, polymer solution or polymer melt in fewer settings \cite{MESpin1, MESpin2} is forced out of a container (emitter) with tiny orifice using a piston, such as a syringe pump. This emitter is connected to the positive terminal of a high voltage DC source ($\sim$ few to 10s of kV) and a flat plate or drum placed at a distance (collector) is connected to the voltage source's negative terminal, thereby setting up a high voltage DC electric field between the emitter and collector. The piston pressure competes with surface tension to generate a polymer jet, whereas the DC electric field accelerates the jet from emitter towards the collector and stretches them into fibers. The fibers are deposited on the collector where they evaporate their solvent to result in the electret fabric. A schematic representation of the electrospinning principle is shown in fig.~\ref{fig1}a. Whereas polymers used in electrocharged fabrics possess embedded charges, the electric field between the emitter and collector aids in orienting dipoles of the polymer melt during droplet stretching, thereby further enhancing the material's electrocharging properties.

The electrospinning method is agnostic to the type of polymer material, with material dependent parameters, e.g. melting temperature and DC field voltage, being easy to adjust for each material once the primary setup is in place. However, the electrospinning process suffers from two disadvantages. Firstly, electrospun polymer throughput scales linearly with the number of orifices or syringes, requiring several syringes in parallel for increased throughput. Secondly, the high voltage DC electric field is expensive and requires additional operational safety features, hence not suitable outside laboratory and industrial settings where the method outlined below is intended to find its primary use.

{\it Melt Blown Process:} The most common method for manufacture of electrocharged filtration fabrics is Melt Blowing \cite{MBlowing1, MBlowing2, MBlowing3}. In this method, jets of molten polymer are generated by injecting it from a conical die, wedged in a gap within an air knife that converge at the die tip; Fig.~\ref{fig1}b shows a cross-sectional view of the cylindrical geometry for a generic Melt Blowing setup. The hot air accelerates and stretches the polymer jet into fiber. Cold air sprayed at the polymer strands as they depart the die tip solidifies them as they land on a conveyor belt or drum \cite{MBlowBook1, MBlowBook2}. Like electrospinning, the melt blown process too can employ a variety of polymers by controlling temperatures of the die (for polymer melting) and hot air jets. Melt blowing offers a higher throughput relative to electrospinning and does not require high DC fields. However, melt blowing setups are inherently suited for large volume manufacturing, requiring a dedicated source of high pressure air at large flow rates, and thus not easily amenable to construction from commonly available parts. The quality of fabric too is ever so sensitive to die geometry and shaping of the high velocity hot air currents, thus requiring time-consuming fine tuning of the process.

\begin{figure*}
\begin{center}
\includegraphics[width= 1.75\columnwidth]{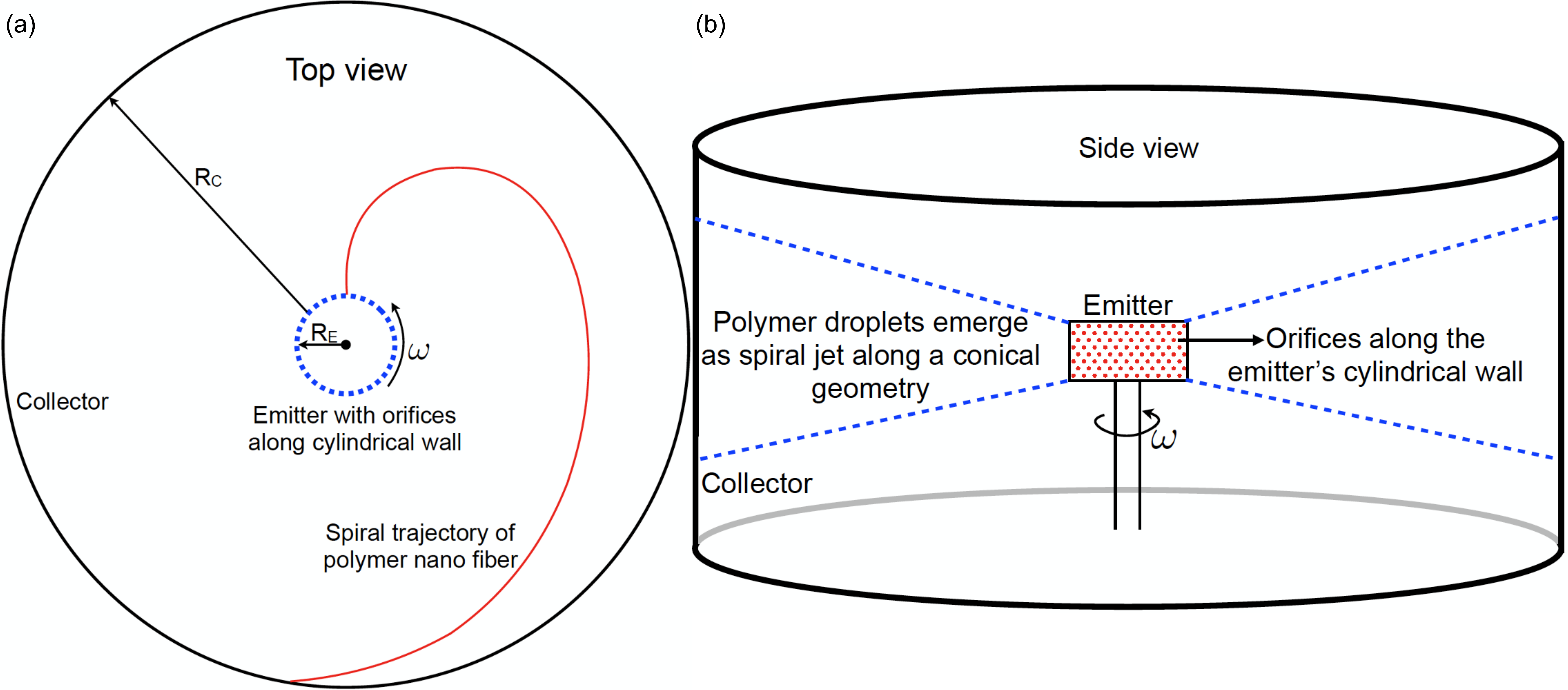}
\end{center}
\caption{(Color online) (a) Top and (b) Side view schematic of Rotary Jet Spinning (CC method).}
\label{fig2}
\end{figure*}

{\it Rotary Jet Spinning:} In recent times, Rotary Jet or Centrifugal Spinning process has emerged as an attractive alternative platform \cite{RJet1, RJet2}. In this method (see schematic in Fig.~\ref{fig2}), a central cylindrical container (emitter) holds the polymer and has several orifices along its wall. The emitter is heated to melt the polymer, but high viscosity prevents it from flowing out of the orifices under static conditions. However, when the emitter is spun at several 1000s of revolutions per minute (rpm), the molten polymer is forced out of the orifices in jets, which are stretched into fibers by centrifugal force generated from fast spinning. A cylindrical drum enclosure (collector) surrounds the emitter and as the stretched polymer jets traverse the distance from emitter to collector in a spiral trajectory as shown in Fig.~\ref{fig2}a, they are cooled and deposited on the collector surface. Rotary jet spinning employs the same forcing scheme to generate as well as to accelerate and stretch polymer jets. It neither requires a high voltage electric field as employed in electrospinning \cite{ES-RSComp} nor hot air jets used in melt blowing. Furthermore, since the emitter wall has several orifices, rotary jet spinning offers higher throughput relative to electrospinning \cite{RSpinReview}. In fact, rotary jet spinning has been long known in regular life through the cotton candy machine and for this reason, it is also called the Cotton Candy (CC) method \cite{CCandy1, CCandy2}. Replacing sugar with polymer and tuning the temperature and emitter rpm offers an easy design one can construct from commonly available parts.

\section{Fabrication Setup}
The ability to construct a low-cost fabrication setup from commonly available parts and materials forms the primary consideration behind the design strategy detailed in this article. For this reason, some design choices were made at the very outset to keep the design rules accessible to the layperson. Firstly, non-woven electret polymer fabrics can also be manufactured by dissolving polymers in suitable organic solvents, as opposed to melting them at high temperatures. However, organic solvents are not commonly available whereas heat sources are universally accessible. For this reason, methods involving liquid polymers dissolved in solvents are not explored here and will form part of a future study. Secondly, as already discussed, CC method offers easy construction from common materials as opposed to electrospinning and melt blowing. The design strategy therefore relies heavily on the CC method (Fig.~\ref{fig2}), with hybrid characteristics that adopt some aspects of electrospinning and melt blowing wherever practicable.

\subsection{Control Parameters}
As discussed earlier, electret fabrication involves forcing in two stages, whose order of magnitude analysis presented here is borrowed from Ref. \cite{RJet2} and the terms defined below are also shown in the schematic in Fig.~\ref{fig2}. The first stage concerns droplet generation in order to initiate a jet by overcoming the capillary force $F_{\sigma} = \sigma r_O$, where $\sigma$ is the surface tension of the polymer melt and $r_O$ is the orifice radius. Hydrostatic pressure being much smaller in magnitude than the centrifugal force $F_{\omega} \sim \rho \omega^2 R_E r_O$ ($\rho$ being polymer melt density, $\omega$ the emitter's angular speed, and $R_E$ emitter radius), balancing the inertial force $\rho \omega^2 R_E r_O^3$ with $F_{\sigma}$ provides the threshold angular speed $\omega_{th}$ for droplet generation and jet initiation:
\begin{equation}
\omega_{th} \sim \sqrt{\frac{\sigma}{r_O^2 R_E \rho}}
\label{omegath}
\end{equation}

The second stage of droplet acceleration or jet elongation concerns a competition between the centrifugal force $F_{\omega}$ and the viscous force $F_{\mu} \sim \mu v/x$ where $v$ is the jet velocity at a distance $x$ from the orifice and $\mu = \sqrt{\sigma R_E \rho}$ is its extensional viscosity since the polymer melt is a viscoelastic fluid and the droplet stretching represents a case of extensional rheology. Applying mass conservation between matter ejected at the orifice with speed $u$ and the elongated jet of radius $r$ arriving at the collector at a distance $R_C$ from emitter, provides the mean fiber radius:
\begin{equation}
r \sim \frac{r_O u^{1/2} \nu^{1/2}}{R_C^{3/2}\omega},
\label{frad}
\end{equation}
where $\nu = \mu/\rho$ is the extensional kinematic viscosity; it is assumed $R_C \gg R_E$.

Whereas equations \ref{omegath} and \ref{frad} resulting from scaling analysis \cite{RJet2} provide appropriate control parameters for the CC method, some of them are invariant. For instance, surface tension $\sigma$ does not vary significantly with temperature $T$ and maybe assumed constant, and since polymers dissolved in solvents by concentration are not explored here, the extensional viscosity $\mu$ varies only with temperature through density $\rho$. Therefore the dominant parameters that control our design are orifice radius $r_O$, emitter radius $R_E$, collector radius $R_C$, angular speed $\omega$ and heating temperature $T$ through which $\rho$ (and $\mu$) is varied. It may then be surmised that the material dependence enters only through the temperature. Armed with these parameters, practical design rules may now be developed.

\subsection{Design Rules}

\subsubsection{Materials}
The material of choice in manufacture of electrocharged filtration fabrics is usually polypropylene (PP) of high molecular weight, but polystyrene (PS) and poly(4-methylpent-1-ene) also possess high electrocharging characteristics. Commonly available materials being the primary goal, PP and PS become the natural materials of choice as they can be easily sourced as raw material, or are commonly available through plastic containers. This has implications for the manufacturing process in that PS being glassy, requires better temperature control, but cooling is less difficult. On the other hand, PP viscosity is nearly insensitive to temperature above its melting point, but does require cooling well below its melting point in order to crystallise.

In using PP, it is known that low molecular weight ($M_W$) isotactic PP ($M_W < 12000$) tends to form brittle fibers that easily break up \cite{Strength}. If working with PP pellets ordered from regular suppliers, high molecular weight isotactic PP is preferred. In the following, fabrics made from isotactic PP were fabricated with material characteristics: $M_W = 250000$, melting temperature $T_M = 160 - 165^{\circ}$C and density $\rho = 900$kg/m$^{3}$; this is in fact the preferred material for manufacture of N95 FR electrocharged layers. Another commonly available polymer used in electrocharged filtration fabrics is polystyrene owing to its high charge retention property, as is known from common experience with styrofoam packaging material which easily sticks to surfaces due to static charge. PS material was employed with material properties molecular weight $M_W = 35000$, glass transition temperature $T_G = 100^{\circ}$C but with a reasonable minimum temperature for processing in the range $T = 123 - 128^{\circ}$C and density $\rho \sim 1060$ kg/m$^3$. 

One can also use discarded PP and PS plastic containers, but care should be taken not to use expanded PS that is generally available in the form of packaging material, but rather regular PS known as General Purpose Polystyrene in industry and used in fabrication of plastic containers through injection moulding. When using discarded PP and PS containers, they were crushed into powder using a commonly available blender. However, PP containers are often manufactured from low crystallinity PP, which results in very dense fabric mesh due to high cohesive properties, and is uncomfortable when breathing \cite{StrucFilt}. Additionally, low crystallinity PP was found not to possess high charge relative to isotactic PP. However, when mixed with PS from discarded containers, the resulting fabric is more compliant, less dense allowing easier respiration, and retains excellent electrocharging characteristics. A mixture of 80\% low crystallinity PP with 20\% PS  (PP-PS blend) gave very good results. It is noted that PP and PS are not miscible, and that by a blend it is merely implied that both powders were molten together.

The melting and extrusion process of PP or PP-PS would likely inactivate most biological material (in particular bacteria, fungi and viruses). As a comparison, autoclaves are typically run at comparable temperatures (160- 190$^{\circ}$C) on a dry cycle for 15-120 minutes. For compatible materials, sterilization is usually done in a wet cycle at 120$^{\circ}$C and fairly high pressure ($\sim$100 kPa) as the steam helps to break open cells and irreversibly denature protein and nucleic acids. Quickly melting polymer powder of contaminated bottles passed through a blender may not result in destruction of 100\% of the infectious properties of potential contaminants. This is of particular concern when using the resulting material for face protection. It is therefore advisable that plastic containers be cleaned in a domestic pressure cooker at high steam for 20 minutes before crushing them in a food processor to turn into powdery material.

\subsubsection{Jet Generation}
{\it Temperature:} Tests on the appropriate range of temperatures were heavily informed by Ref.~\cite{CCandy2}. Although the transition temperature for PS is $T_T = 123 - 128^{\circ}$C and that of PP is $T_T = 160 - 165^{\circ}$C, they do not readily flow at these temperatures but merely soften into highly viscous fluids. Whether working with pure PP or PP-PS, heating the emitter to higher temperatures in the range $T = 175 - 200^{\circ}$C reduces the viscosity, but the jets tend to break up during extension and do not result in fibers. We found that fibers were indeed generated in the temperature range $T = 200 - 250^{\circ}$ but resulted in beaded structures. Increasing the emitter angular speed $\omega$ did result in continuous fibers in accord with published literature \cite{RJet2}. However, temperatures $T > 280^{\circ}$ reduced viscosity sufficiently to give continuous fibers even at lower angular velocities. Note that Eq.~\ref{frad} shows mean fiber radius $r \sim \nu^{1/2}$ and inversely proportional to $\omega$. It is therefore desirable to tune the heating temperature $T$ to match the maximum rpm achievable by the motor employed to spin the emitter as discussed below.

Small commercial cotton candy machines usually employ electrical heating elements whereas larger ones are gas fired. Irrespective of the heating method, most cotton candy machines operate at temperatures around $T = 160 - 175^{\circ}$C, which falls below the temperature range desirable for generating jets of pure PP or PP-PS ($T = 280 - 340^{\circ}$). If working with commercial machines, {tweaking the heating elements to achieve the desirable temperature range is suggested}. If building one's own machine, developing one's own electrical heating element is desirable if one has working knowledge, since electrical heating elements provide precise control. A simpler alternative is to use a gas torch and tweak the torch flame and its distance from emitter by checking emitter temperature with an ordinary thermometer used for home baking.

{\it Emitter Motor:} The choice of emitter motor for CC method is dictated by Eq.~\ref{omegath} as it sets the lower bound on the revolutions per minute (rpm) to overcome $\omega_{th}$. Taking surface tension of PP to be $\sigma_{PP} \sim 20 \times 10^{-3}$ N/m, $\rho = 900$ kg/m$^3$, $r_O$ of order 100s of microns ($\sim 10^{-4}$ m), and $R_E$ of order few cm ($\sim 10^{-2}$ m) yields $\omega_{th} \sim$ 470 s$^{-1}$ or 4500 rpm. In practice however, fibers emerge at around 2500 rpm, which is easily achieved with most commercial cotton candy machines as they operate between 3000 - 4500 rpm. If constructing one's own machine, high rpm DC motors capable of going up to 15000 rpm are suggested. A simple alternative is to repurpose an electrical drill to spin the emitter as they commonly achieve up to 4500 rpm or a Dremel drill with variable adjustable speeds in the range 3000 - 37000 rpm. Drills can be easily connected to an emitter container with a suitably threaded steel rod, nut, and washer. Two simple alternatives exist in the event a high rpm motor is unavailable, as presented below when discussing emitter geometry.

\begin{figure*}
\begin{center}
\includegraphics[width=1.5\columnwidth]{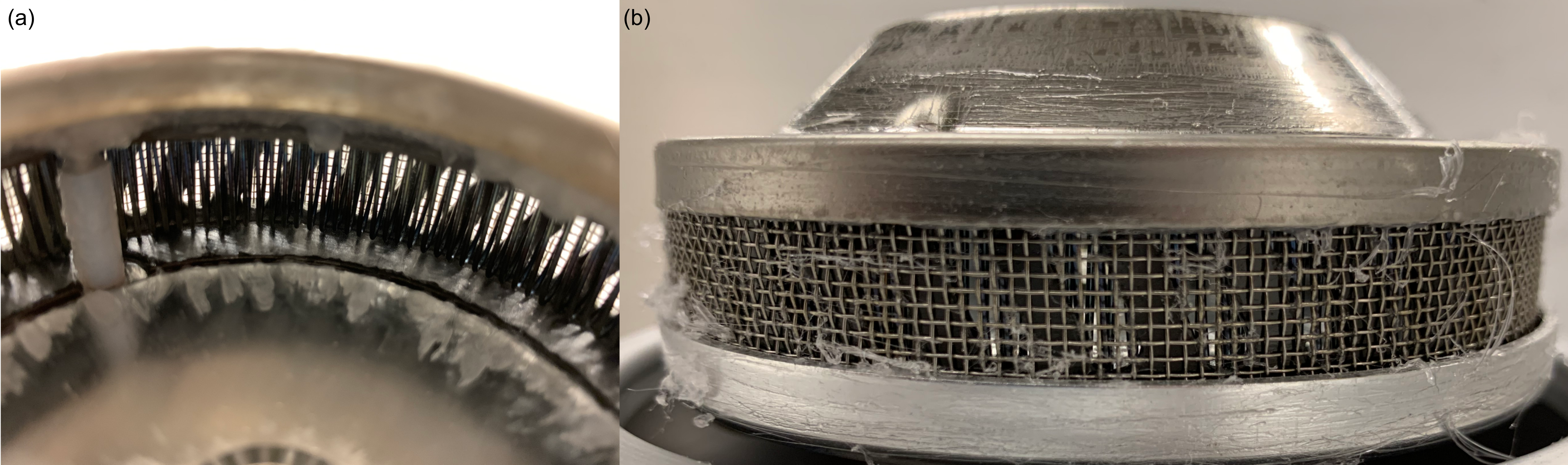}
\end{center}
\caption{(Color online) Commercial candy machine emitters usually come with ((a) vertical or diagonal slats or (b) wire mesh for orifices.}
\label{fig3}
\end{figure*}

{\it Emitter:} The emitter follows a standard cylindrical geometry of radius $R_E \sim 5$ cm (0.05 m) and height of roughly 0.1 m. The side wall of the cylindrical emitter is dotted with several tiny orifices through which the polymer jets are forced out when the emitter is spun. Since orifice radius $r_O$ and emitter radius $R_E$ are the system design parameters entering Eq.~\ref{omegath} ($r_O$ also enters Eq.~\ref{frad}), they should be chosen to match the material parameters $\sigma$ and $\rho$ in order to obtain a comfortable $\omega_{th}$ within the operational rpm range of one's chosen emitter motor. Smaller the orifice radius $r_O$, higher is the required $\omega_{th}$ as they're inversely proportional, thus setting the upper bound on choice of $r_O$. However, the mean fiber radius $r$ scales linearly with $r_O$, and constrains the upper bound through Eq.~\ref{frad}. If one has access to high rpm motor, then it is advisable to go to lower $r_O$ of order few 100 microns.

Commerical cotton candy machines either use an emitter mesh with large orifice dimensions of order 1-2 mm or vertical/diagonal slats of width $\sim 1$ mm, see fig.~\ref{fig3} for an exemplar. Replacing such emitters with home-built emitter cylinders of lower orifice radii is suggested. Alternatively, if constructing one's own fabrication system, an easy way to construct the emitter is using a soda or beer can as they come readily manufactured to the right radial dimension and aluminum is an excellent thermal conductor. Cutting or shearing a soda or beer can in half and using the bottom half gives a readymade emitter. Folding the top open edge along the wall perimeter and drilling holes of few 100 micron radius along the emitter wall provides satisfactory results. For instance, the present study used drill bits of gauge 87 to obtain $r_O = 0.254$ mm with a Dremel drill motor connected to bottom half of a soda can with a nut and washer to drive the emitter, which gave very satisfactory results.

In the event, one does not have a high rpm motor, a hybrid design combining rotary jet spinning (CC method) with melt blowing works just as well. Recognizing that Eq.~\ref{omegath} results from a balance between surface tension and inertial forces because hydrostatic pressure is low, one could consider sealing the top of the emitter and pump compressed air at roughly 0.2 - 0.3 MPa, while spinning the emitter at lower rpm to overcome the surface tension force to generate polymer jets. This is simply the melt blowing principle in disguise, and it shifts the inertial force by a DC offset value proportional to the pumping pressure divided by emitter wall's surface area $\pi R_E^2 h$ ($h$ is emitter height), thus bringing down the $\omega_{th}$ value. Of course, as a consequence, the lower $\omega$ would lead to larger mean fiber radius since $r \sim 1/\omega$ but this can be overcome by either decreasing $r_O$ and/or increasing $R_C$, the latter usually being easier to accomplish.

\subsubsection{Fluid Acceleration and Stretching} Fluid acceleration and stretching is the stage at which the fibers are formed. Once the jets are generated, they're flung radially outwards and execute a spiral trajectory as shown in Fig.~\ref{fig2}b before they're deposited on the collector surface. From Eq.~\ref{frad}, the mean fiber radius $r \sim R_C^{-3/2}$ scales linearly with $r_O$ and inversely with $\omega$, which together form the system design parameters controlling fiber radius. It is therefore advantageous to have as large a collector radius to achieve fibers of ever smaller radii. These system design parameters are complemented by the polymer's kinematic extensional viscosity $\nu = \mu/\rho$ which scales as $r \sim \nu^{1/2}$. For this reason, emitter temperatures in the range $T = 280 - 340^{\circ}$C are suggested so that the viscosity is low enough for the polymer melt to flow, but still high enough to result in continuous, fibers of lower radii. The droplet ejection velocity $u$ is not a controllable parameter since it is a resultant of the competing (viscous, surface tension, and centrifugal) forces.

Typical range of collector radii for commercial candy machines fall in the range $R_C \sim 0.1 - 0.3$ m. Naturally, larger the $R_C$, the finer the fiber radii generated, by virtue of Eq.~\ref{frad}. If constructing one's own fabrication system, one has the freedom to set $R_C$ to larger radii. We have found $R_C \sim 0.25 - 0.4$ m worked best as they gave sufficient distance for fibers to stretch and cool as they traverse from emitter to collector. Furthermore, if one's emitter motor is unable to achieve high rpm, we suggest a simple hybrid solution borrowed from electrospinning to circumvent the problem. It was observed that at low rpm $\sim 3000 - 4000$ rpm, and $R_C \sim 0.3$ m, connecting a car battery or laboratory power supply at 12 - 24V DC (negative terminal to emitter and positive terminal to collector with insulation between the two) generates a DC electric field to add sufficient droplet acceleration to stretch the fibers to desired radii. Irrespective of acceleration needs, the electric field also plays an indirect role in enhancing the fabric's surface charge characteristics, as discussed later.

\subsubsection{Electrocharging}
Electrets -- dielectric materials which exhibit an external electric field in the absence of an applied field -- can be broadly classified in two varieties, {\it viz.} space-charge or real-charge electrets and dipolar or oriented-dipole electrets \cite{Electrets}. Real-charge electrets possess an injected or embedded excess charge (of one or both polarities) within the dielectric volume or at the surface; they are usually manufactured via direct charge injection into the dielectric material. Dipolar electrets are formed by dipole orientation (polar groups) within the dielectric material and are usually formed, or rather polarized, by applying an electric field to the material either at an ambient temperature or by heating the material above its transition temperature $T_T$ where dipoles become mobile. An alternative method is to employ charge injection techniques on dipolar electrets where the embedded charge causes dipole reorientation. The fundamental limitation to all electret-charging methods is dielectric breakdown of the polymer material or external medium, which depends on the dielectric strength of a given material (typically of order few MV/cm for polymers with charge densities of a few tenths of  $\mu$C.cm$^{-2}$). The primary methods for electric charging of electrets include, Triboelectric charging, Thermal charging, Isothermal charging, Electron and ion beam charging and Photoelectret charging.

\begin{figure*}
\begin{center}
\includegraphics[width= 1.95\columnwidth]{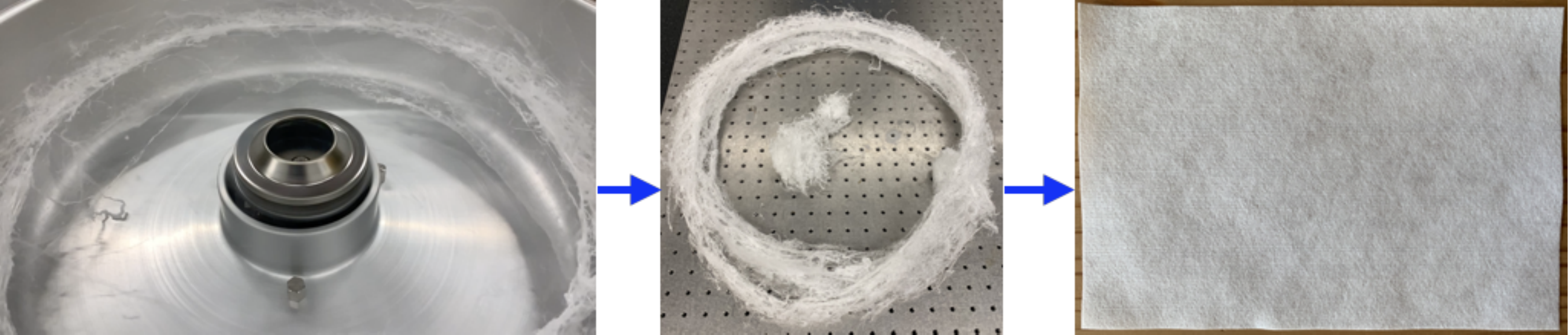}
\end{center}
\caption{(Color online) Fabrication flow process: Fibers generated in a modified cotton candy machine (see movie M1.mov in supplementary material for brief movie of the process) could be scooped out as a concentric fiber fabric, which was sandwiched and cut into individual sheets. The sheets were then subjected to isothermal charging with an air ionizer 1 cm from the fabric for a 10 minute duration to improve the fabric's static charge characteristics.}
\label{fig4}
\end{figure*}

As the name suggests, triboelectric charging occurs from charge transfer due to frictional contact between dissimilar dielectric materials. Not only is this method very unreliable as it requires intimate contact between the two surfaces being charged, the present study does not employ two different materials, hence triboelectric charging does not apply here. Thermal charging involves application of an external electric field to the polymer at elevated temperatures, as occurs with the electrospinning process. Even though PP and PP-PS materials used here fall under the real-charge electret classification, application of a weak (12-24 V) DC electric field as described in the previous subsection did enhance surface charge characteristics as shown later. Electron and ion beam charging involves low energy secondary electron cascade resulting from scattering of the primary beam within the dielectric bulk and is not generally efficient for non-woven polymer fabrics of the kind explored here. On the other hand high energy electron beams cause chemical damage to most dielectric materials, hence not a suitable process for present needs. The photoelectret process only applies to charging of photoactive polymers and not relevant to the current study.

The best results were obtained by Isothermal charging \cite{coronachg1, coronachg2}. In this method the polymer fabric is placed between two electrodes maintained at high electric fields (typically kilovolts and 100s of $\mu$A current). Although this method may seem complicated due to requirement of high voltages, it is in fact the easiest to achieve \cite{CoronaPP1, CoronaPP2}. Ionizing air purifiers used in homes and offices operate on the isothermal charging principle, where they apply high voltage to ionize or electrically charge air molecules to attract charged dust particles, bacteria, and viruses. Air ionizers come in two varieties -- those generating negative ions (anions) and electrostatic discharge (ESD) ionizers (balanced ion generators). ESD ionizers should not be used for iosthermal charging because not only do they not impart charge, they in fact neutralize existing charges on surfaces. The relevant ionizers suitable for electrocharging fabrics are anion generators. The present study used an ionizing air purifier for home settings (Model NIP-6E from Mystic Marvels LLC) operating at 9kV and 160 $\mu$A. Exposing the fabricated polymer fabric to isothermal charging for 10 minutes at a distance of 1 cm (0.1 m) substantially enhanced their surface charge characteristics. Details of the charge characterization are presented in the next section.

\subsection{Fabrication Process}
\subsubsection{Setup and Material}
Having outlined the primary control parameters for the CC method and various ways to optimize them in the design process, the basic fabrication process is now explained. The dimensions and operating ranges naturally vary by user; the fabrication setup employed in this study had emitter radius $R_E = 0.055$ m, collector radius $R_C = 0.35$ m, and orifice radius $r_O = 254~\mu$m ($2.54 \times 10^{-4}$ m). The emitter heating element could be adjusted to achieve temperatures up to $T = 400^{\circ}$C, but the operating temperature in this study was $T = 285^{\circ}$C for PP-PS and $T = 300^{\circ}$C for PP. The emitter motor could achieve a maximum rpm of 35000, but the emitter was run at 10000 - 12000 rpm, capable of yielding a throughput of approximately 0.65 kg of fibers per hour.

The basic process steps are outlined in Fig.~\ref{fig4}, also see movie M1.mov in supplementary material for a brief video of fiber generation process. Each fabrication run used 12 grams of polymer as input, and the resulting fibers were available in less than a minute. The fibers scooped out from the collector wall had a concentric circular profile, which were then sandwiched between two flat and clean surfaces at high pressure, this study employed 1 cm thick float glass plates, resulting in thin fabric sheets of 0.2 - 0.3 mm thickness. The values vary with the collector dimensions, quantity of material used per fabrication run, pressure applied to generate the final fabric sheet and are best worked out through trial-and-error by each end user. Characterisation details for fabrics are provided in the next section.

\subsubsection{Face Masks}
\begin{figure*}
\begin{center}
\includegraphics[width= 1.9\columnwidth]{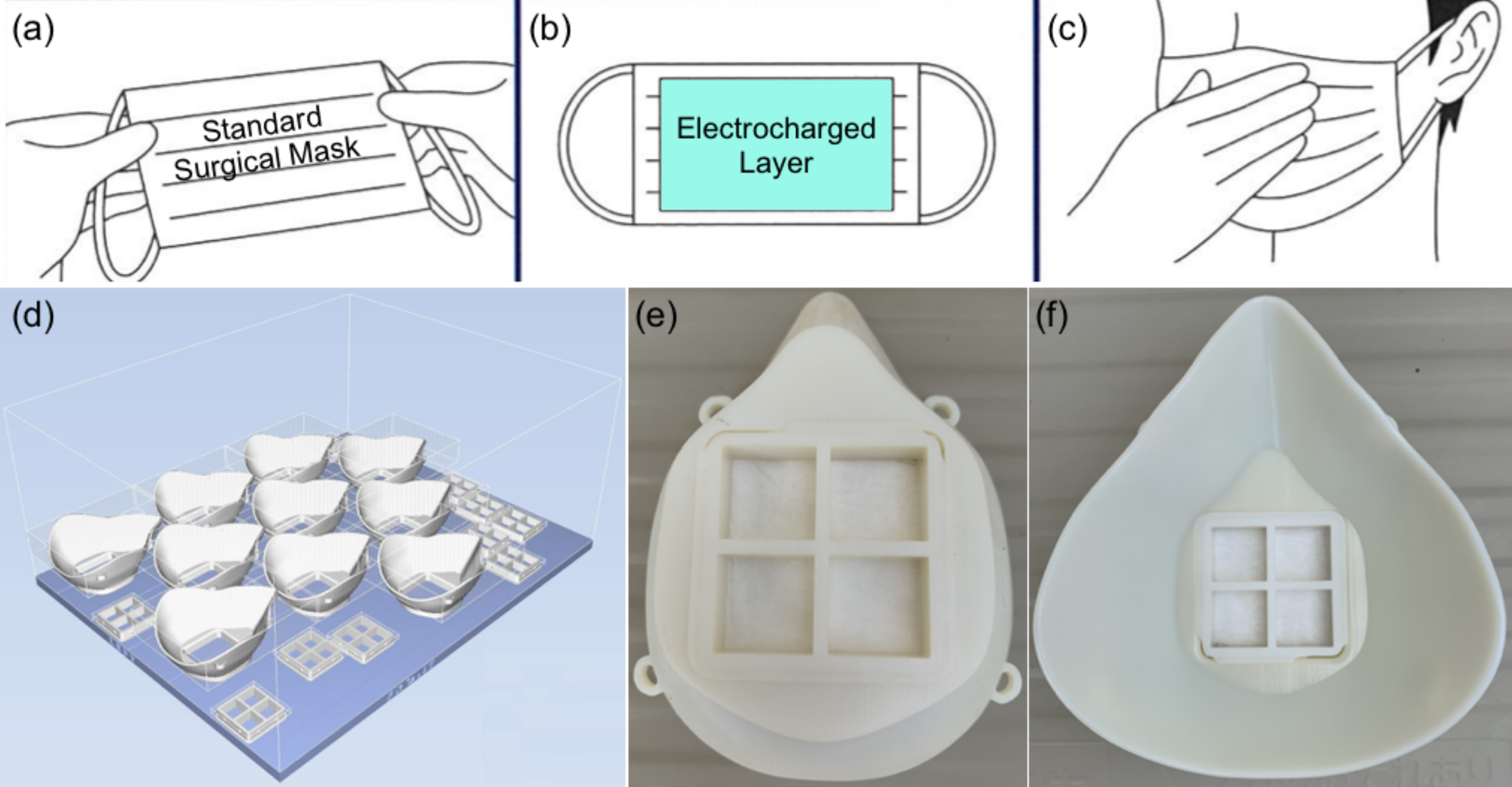}
\end{center}
\caption{(Color online) (a) Standard surgical masks (typical dimensions 9 cm $\times$ 16 cm) were (b) reinforced with an electrocharged filtration layer (5 cm $\times$ 10 cm) fabricated in-house and attached to mask's inner surface to improve filtration efficiency upon (c) wearing the mask. Montana Mask holders with tight facial fit with (d) placement design for Object 500 3D printer and the (e) front and (f) back view of finished Montana Mask, to hold square (5 cm $\times$ 5 cm) patches of electrocharged filtration fabric (3 layers per mask) to achieve N95 filtration quality.}
\label{fig5}
\end{figure*}

Two approaches were followed to turn the fabricated material into face masks. Lacking in-house capability to stitch fabric layers into masks, in the first approach a sheet of the electrocharged fabric manufactured by the CC method was added to the inner surface of surgical and PM2.5 face masks available commercially as shown in Fig.~\ref{fig5} (a-c). Three different sets of tests were performed on these masks. The first, a control test, was performed on the surgical masks in the condition they were procured to measure their baseline filtration quality. In the second test, a sheet of the manufactured fabric was added after intentionally bleeding the sheet of its electrostatic charge with a static eliminator. The reason for this second test was to discriminate between filtration quality arising purely from the presence of an additional porous layer without electrocharged filtration capability. The final test was performed on a surgical mask with an electrocharged fabric layer added to its inner surface. In all three cases, the static electricity on the masks before and after filtration tests were measured with a non-contact electrostatic potentiometer (KSS-2000 Digital Electrostatic Potentiometer, Manufacturer: Kasuga Denki Inc.). Whereas these filtration tests did show improvement in filtration quality due to addition of the electrocharged fabric layer, as is well known, the lack of tight facial fit left gaps through which particles could easily pass unfiltered.

In the second approach, face mask holders were 3D printed from an open-source design known as the Montana Mask \cite{MMask} as shown in Fig.~\ref{fig5}(d-f)) to overcome the deficiency of standard surgical masks in providing a tight facial fit. The Stereolithography (STL) format design files for the Montana Mask are publicly available for download from Ref.\cite{MMask}. One of the design features that allows N95 FRs achieve superior filtration efficiency of $\ge 95\%$ is their ability to provide tight facial fit and prevent air from leaking through the interstitial gap between the mask and skin during respiration. Mask holders (fig.~\ref{fig5}d) were manufactured on an Objet 500 3D printer from the open source STL files \cite{MMask}. The filter holder shown in fig.~\ref{fig5}e-f (front and back views, respectively) with a square grid could hold a square patch roughly 5 cm $\times$ 5 cm in dimensions. Adding up to three electrocharged filtration layers resulted in the desired N95 filtration quality, but respiration became difficult with 5 layers because the fabricated layers were denser than the layers present in commercial N95 FRs. These  3D printed Montana masks were therefore limited to 4 electrocharged filtration layers for optimal respiratory comfort without sacrificing filtration quality. Table~\ref{table1} lists the mean densities (g.cm$^{-3}$) obtained by dividing the measured weight by the dimensions of the samples for commercial fabrics and ones manufactured in-house. The standard deviation is quoted over measurements for ten samples of each material, except for the N95 FR for which only one sample was used. All values are rounded off to second decimal place, and the raw data is available at Ref.~\cite{BandiN95-Data1}. The variability is higher in fabrics manufactured in-house, and that is to be expected strict quality control processes possible in industrial methods were not permissible in the in-house fabrication method.

\begin{table*}
\caption{Electrocharged fabric densities, data available at Ref.~\cite{BandiN95-Data1}}
\begin{tabular}{cc}
\hline
Material & Density (g.cm$^{-3}$): Mean $\pm$ Stdev.\\
\hline
N95 FR & 0.45\\
Surgical mask & 0.38 $\pm$ 0.03\\
Isotactic PP & 0.73 $\pm$ 0.09\\
PP-PS & 0.63 $\pm$ 0.09\\
\hline
\end{tabular}
\label{table1}
\end{table*}

\begin{figure*}
\begin{center}
\includegraphics[width= 1.95\columnwidth]{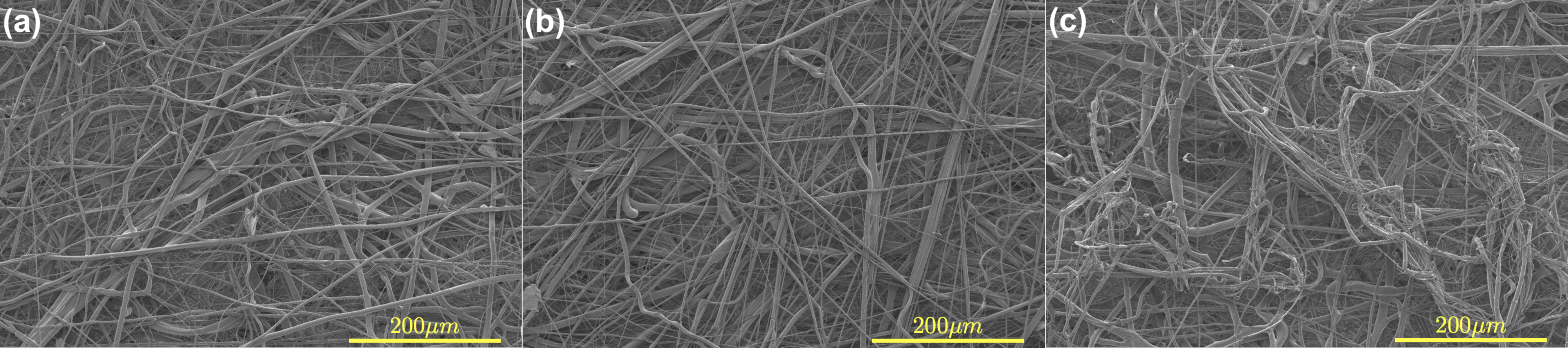}
\end{center}
\caption{(Color online) Scanning electron micrographs for electrocharged fabrics from (a) a commercial N95 FR, (b) isotactic Polypropylene (PP) fabricated in-house, and (c) Low crystallinity Polypropylene - Polystyrene (PP-PS) blend fabricated in-house. High resolution images available at Ref.~\cite{BandiN95-Data1}}
\label{fig6}
\end{figure*}

\section{Characterisation}
Following details of the design rules and fabrication process, characterisation studies performed on the fabricated electrocharged material and resulting face masks are now presented for the structural, charge retention, and filtration properties of fabrics and masks developed in-house. Recognizing that knowledge of their properties were not meaningful by themselves, the same tests were performed on a NIOSH-certified commercial N95 FR to serve as the benchmark against which to compare the quality of our fabrics and masks.

\subsection{Structure}
Structural properties of the electrocharged filtration fabrics were studied using scanning electron microscopy (SEM). As a preparatory step, platinum-palladium sputter coating deposition was performed on the fabric sample surfaces for SEM visualization, followed by interrogation under a scanning electron microscope (Quanta 250 FEG, Manufacturer: FEI Thermo Fisher) at 2 kV acceleration voltage. 

Figure~\ref{fig6} presents scanning electron micrographs for electrocharged fabrics. In qualitative terms, the N95 electrocharged fabric layer (fig.~\ref{fig6}a) seems structurally similar to PP (fig.~\ref{fig6}b) and PP-PS (fig.~\ref{fig6}c) fabrics in terms of obtaining a heterogeneous, non-woven fabric of enmeshed fibers. This heterogeneous structure results from fluctuations in fiber trajectories arising from the individual forcing conditions, {\it viz.} acceleration under DC electric field in electrospinning, hot air jet in case of melt blowing, and centrifugal forcing for the CC method. The qualitative similarity in fabric heterogeneity obtained by CC method (fig.~\ref{fig6}b and c) relative to N95 electrocharged fabric (fig.~\ref{fig6}a) presumably manufactured via melt blowing was therefore very encouraging.

A cursory inspection suggests the commercial N95 FR fabric's fibers were slightly more tortuous relative to PP fabric, and less tortuous than PP-PS fibers. To understand the qualitative difference in fiber tortuosity, it is recalled that  PP melt forms crystalline fibers whereas PS is a glass. PP fibers are therefore expected to result in linear, crystalline fibers relative to PS fibers which freeze into tortuous structures as the molten fluid undergoes glass transition under cooling and viscosity abruptly shoots up. Within PP, isotactic PP is more crystalline than low crystallinity PP obtained from discarded plastic containers. It is therefore important to ascertain how degree of crystallinity affects tortuosity or other structural characteristics. Fiber tortuosity is expected to impact the fabric's porosity. Though porosity could not be measured directly, filtration tests did present a small measurable difference between isotactic PP and PP-PS fabrics, which may be attributed to charge characteristics rather than porosity.

\begin{figure*}
\begin{center}
\includegraphics[width= 1.5\columnwidth]{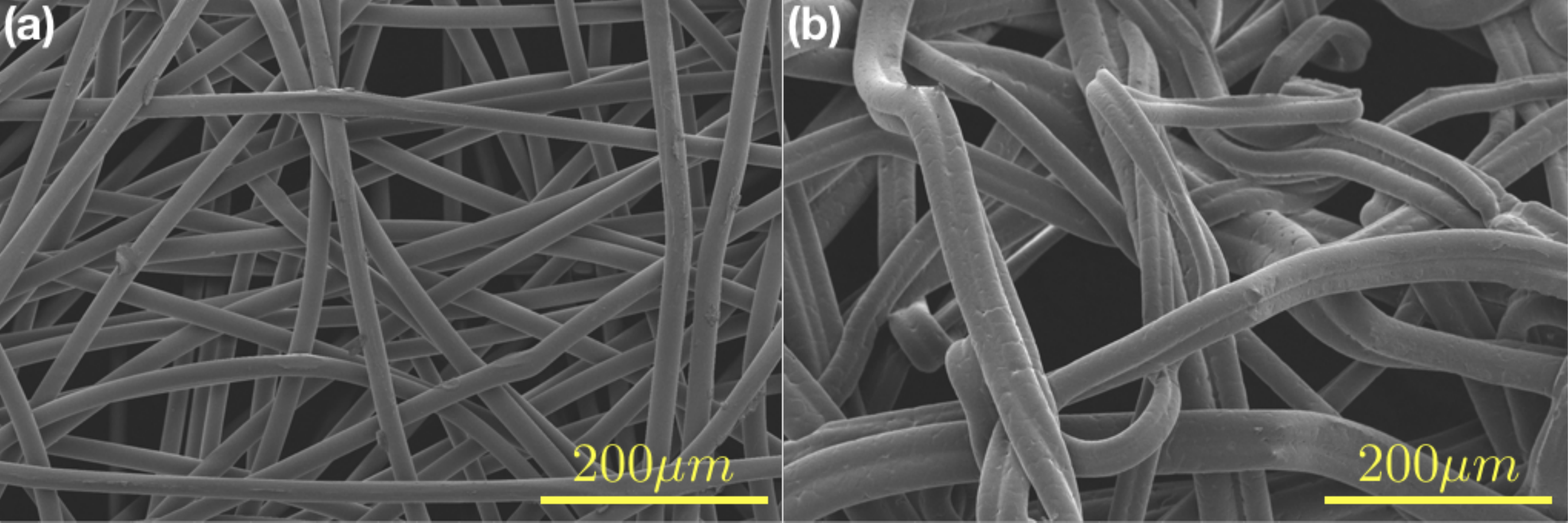}
\end{center}
\caption{(Color online) Scanning electron micrographs for electrocharged fabrics manufactured in-house with (a) Low crystallinity Polypropylene and (b) Polystyrene. High resolution images available at Ref.~\cite{BandiN95-Data1}}
\label{fig7}
\end{figure*}

Figure~\ref{fig7} presents scanning electron micrographs for low crystallinity PP (fig.~\ref{fig7}a) and PS (fig.~\ref{fig7}b) fabrics. Firstly, a comparison of isotactic PP (fig.~\ref{fig6}b) and low crystallinity PP (fig.~\ref{fig7}a) fabrics manufactured under similar conditions shows both fibers are relatively linear. However,  low crystallinity PP fibers possess a more uniform and thicker radius and its fibers are more linear than isotactic PP fibers. When manufactured under similar conditions, mean fiber radii for low crystalline PP were almost twice that of isotactic PP (see Table~\ref{table2}). This implies, the presence of crystalline order, be it low or high, is sufficient to obtain relatively linear fibers whereas the degree of crystallinity determines the average fiber radius under identical fabrication conditions (temperature and rpm).

A comparison of scanning electron micrographs for low crystallinity PP (fig.~\ref{fig7}a) and PS (fig.~\ref{fig7}b) shows PS fibers are far more tortuous owing to their glassy nature, albeit with less variability in fiber radius (see Table~\ref{table2}). Combining the two materials in different proportions helps control the fiber tortuosity, and therefore the porosity, as done in this study which used 80\% low crystalline PP with 20\% PS by weight (fig~\ref{fig6}c). Commercial N95 FR fabrics most likely use a proprietary mix of polymers to control porosity and charge retention, but the specific materials and their percentages are not available in the public domain. The mean and standard deviation of the fiber radii obtained from image analysis using ImageJ open source software of the scanning electron micrographs for all material combinations investigated are listed in Table~\ref{table2}. High resolution images for scanning electron micrographs presented in fig.~\ref{fig7} and fig.~\ref{fig8} are available at Ref.~\cite{BandiN95-Data1}.
 
\begin{table*}
\caption{Electrocharged fabric fiber diameters}
\begin{tabular}{cc}
\hline
Material & Radius: Mean $\pm$ Stdev ($\mu$m)\\
\hline
N95 FR & 4.1 $\pm$ 4.7\\
Isotactic PP & 4.54 $\pm$ 6.2\\
PP-PS & 4.9 $\pm$ 5.1\\
Low crystallinity PP &  9.63 $\pm$ 2.4\\
PS & 18.5 $\pm$ 2.8\\
\hline
\end{tabular}
\label{table2}
\end{table*}

\subsection{Filtration}
\begin{figure}
\begin{center}
\includegraphics[width=\columnwidth]{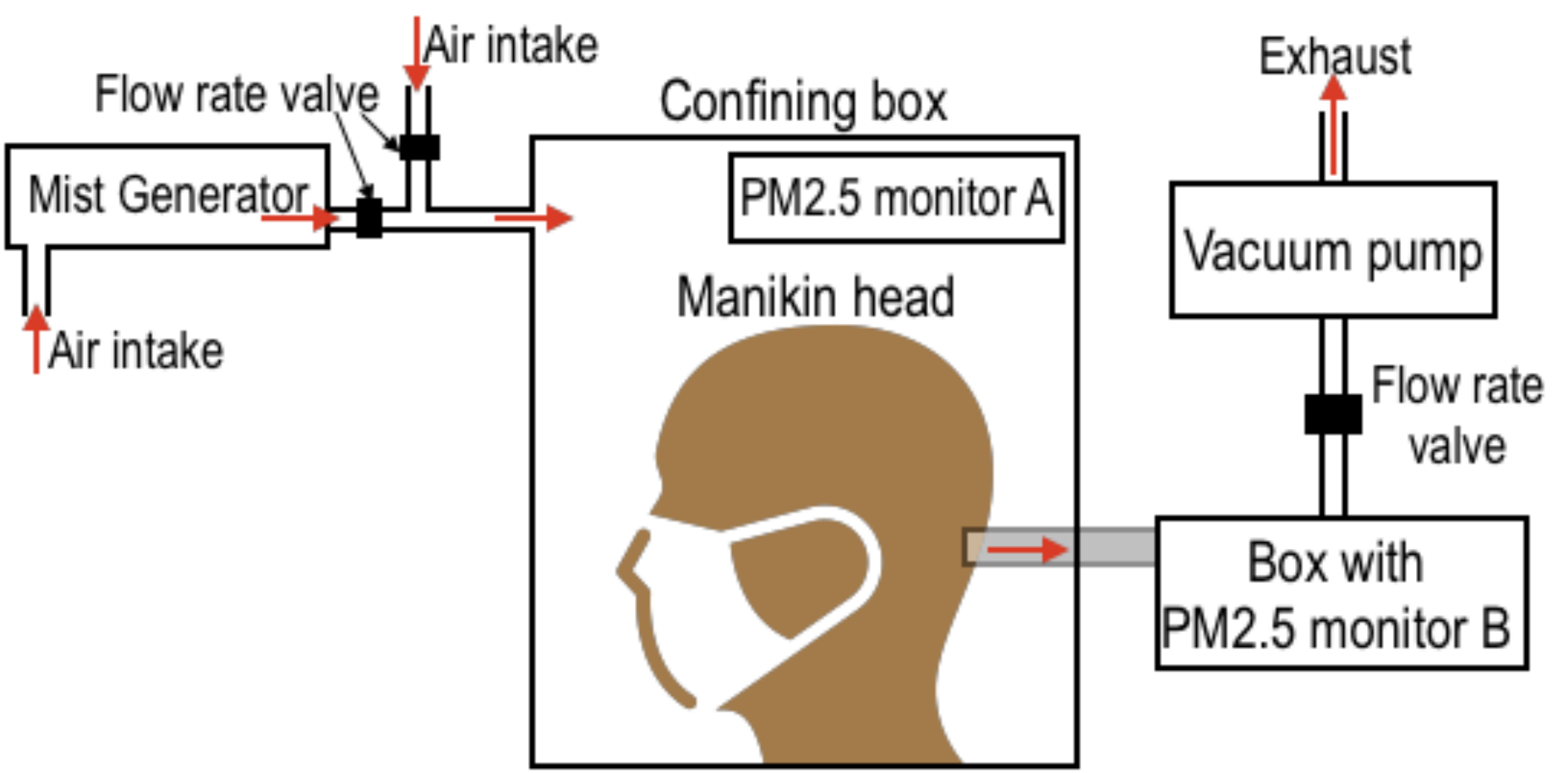}
\end{center}
\caption{(Color online) Filtration test setup schematic.}
\label{fig8}
\end{figure}

{\it Setup:} Filtration tests for face mask certification are usually performed on specialized equipment such as the Portacount Respirator Fit-Tester and MITA 8120, both from TSI Inc. or AccuFIT 9000 from Accutec-IHS Inc. Lacking access to such special testing system and its non-affordability, a filtration testing system was designed in-house as shown in fig.~\ref{fig8}. A manikin head used in retail store fronts was drilled with a hole from its mouth to the back of its head. The face mask under test was then mounted onto the manikin head's face and placed in a confining box. An inexpensive piezeoelectric atomizer (APGTEK Aluminum Mist Maker) usually employed in home decoration was submerged in Sodium Chloride solution (5\% by weight NaCl in de-ionized water) to generate aerosol particles. The generated mist was exposed to negative ion air purifier to charge the aerosol particles for some of the tests. The mist could pass through a pipe with a second connecting pipe open to ambient air as shown in fig.~\ref{fig8} and both pipes had valves to help control the total aerosol concentration in the air entering the confining box.

A portable PM2.5 air quality monitor (Manufacturer: Dienmern) used normally for home and office air quality monitoring was placed in the confining box (Monitor A in fig.~\ref{fig8}) to measure the particle concentration within the box. By reading this PM2.5 monitor, the two inlet valves were adjusted for aerosol mist and ambient air to control aerosol concentration in the confining box. The back of the manikin head was connected to a pipe which exited the confining box and terminated in a box containing a second PM2.5 air quality monitor (Monitor B in fig.~\ref{fig8}). This monitor gave reading of particles that had passed through the fabric and manikin mouth and allowed measurement of filtration quality. This box containing the second PM2.5 monitor was, in turn connected to a vacuum pump as shown in fig.~\ref{fig8}. When the vacuum pump was turned on, a suction pressure was felt in the confining box and aerosol particles mixed with ambient air were sucked into the confining box and passed through the face mask to enter the drilled hole in the manikin head and exited the confining box. By controlling the vacuum pump valve, one was able to simulate flow rates for normal (30 liters per minute) and high (85 liters per minute) respiration rates.

The general ambient conditions under which all filtration tests were conducted are now specified before explaining some of the design shortcomings of the home-designed filtration test setup. The lab temperature during all tests was maintained at 23 $^{\circ}$C $\pm$ 2$^{\circ}$C and relative humidity of 43\%. The relative humidity within the filtration test setup's confining box was however higher due to aerosol presence at 58\%. The area of fabric samples used for surgical masks (SM) was 9 cm $\times$ 16 cm with the electrocharged filtration layer occupying an area of 5 cm $\times$ 10 cm, where as the fabric area in the 3D printed montana mask was 5 cm $\times$ 5 cm. The aerosol number concentration at 30 L/min flow rate was $1.7 \times 10^{8}$ particles per cm$^{3}$ and at 85 L/min flow rate was $1.25 \times 10^{8}$ particles per cm$^{3}$, providing high enough aerosol concentration. Two tests were performed to study time to failure, i.e. the total duration from test commencement after which filtration started to deteriorate and varied between 14.5 hours (for PP fabrics) to 17 hours (for PP-PS blend). As a comparison commercial N95 FRs failed after 22 hours. NIOSH certification standard for N95 FRs requires use of neutrally charged sodium chloride solution (10\% NaCl in water). However the filtration tests in the current study had a dual role, firstly to demonstrate the efficacy of electrocharged filtration, and secondly to test the filtration efficiency itself. To demonstrate the efficacy of electrocharged filtration, independent tests were conducted with aerosol particles that were charged and charge neutralized for N95 FR and the 3D printed Montana Mask design.

{\it Design shortcomings: }It is emphasized that this filtration test setup does not conform to some of the stringent testing specifications employed in face piece respirator certification. For instance, NIOSH 42 CFR Part 84 standard for N95 FRs requires filter performance of $\ge 95\%$ with NaCl test agent at 85 liters per minute flow rate and inhalation resistance (maximum pressure drop across mask) of $\le 343$ Pa and exhalation resistance of $\le 245$ Pa \cite{3M}. The filtration test setup developed in-house had no means to measure the pressure drop nor could one simulate the oscillatory respiratory air flow from inhalation and exhalation and the scheme could only generate steady suction flow.

An important quantity in face mask filtration quality testing is the Most Penetrating Particle Size (MPPS); MPPS for N95 FRs is 300 nm or 0.3 $\mu$m. An important shortcoming of the filtration test setup detailed here is that it cannot provide a size distribution of aerosol particles generated by the relatively inexpensive piezoelectric atomizer. Secondly, commercial respirator testing systems use laser-based particle counter sensors that are sensitive to detection of particles below the MPPS value down to about 100 nm. The PM2.5 air quality monitors, as their name suggests, are rated for measuring particles as small as 2.5 $\mu$m. Be that as it may, PM2.5 monitors also employ the same laser-based particle counter sensors and do hold the capability to detect particles down to 0.3 $\mu$. Although a proper verification was not possible, it is reasonable to assume the PM2.5 monitors could detect particles at least down to 0.3 $\mu$m diameter.

\begin{figure}
\begin{center}
\includegraphics[width=\columnwidth]{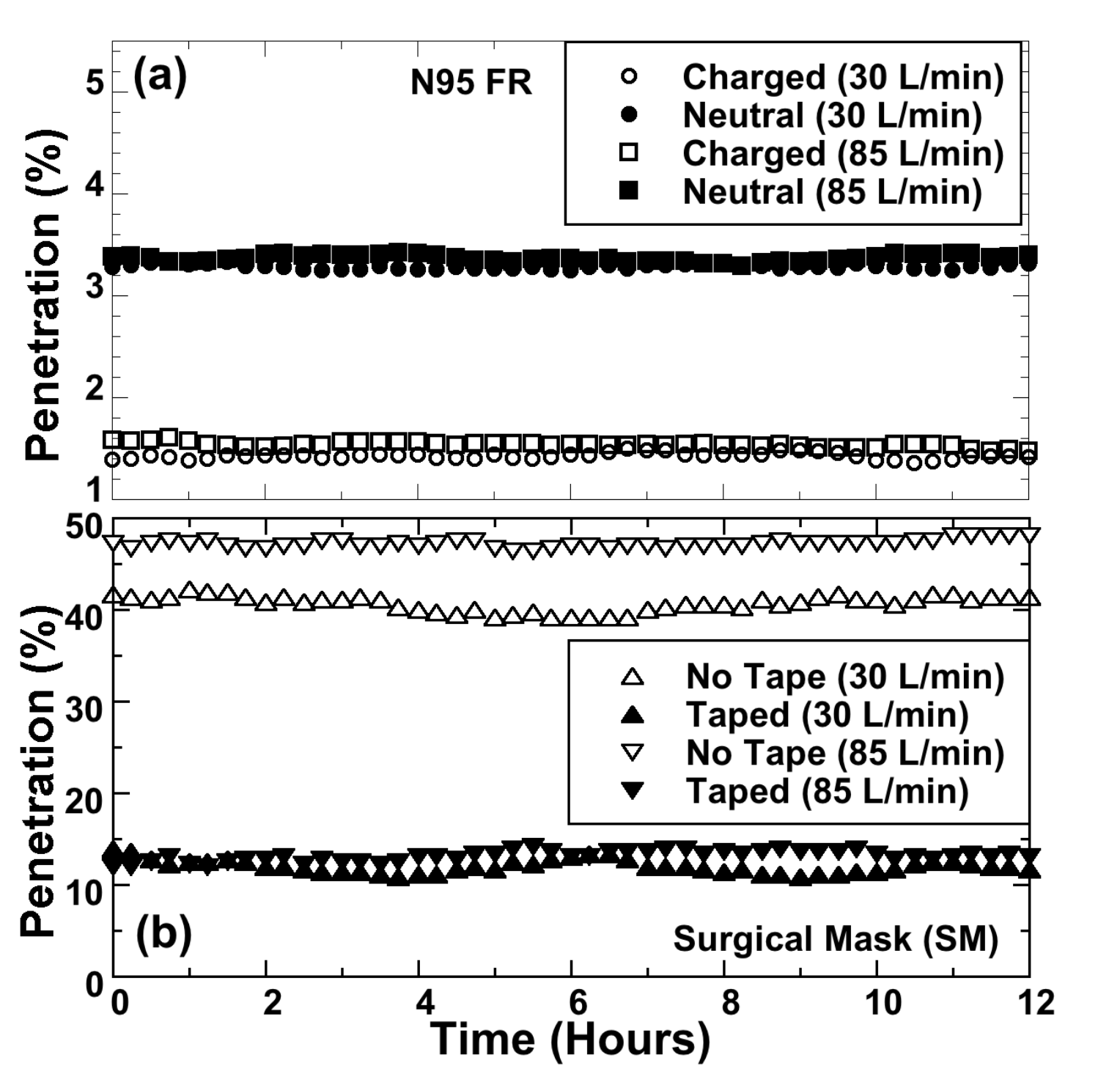}
\end{center}
\caption{Filtration test results showing penetration P(t) (\%) versus time (hours) over 12 hour duration for (a) N95 FR with charged (open black symbols) and neutral (filled black symbols) aerosols at 30 L/min (circles) and 85 L/min (squares) flow rate and (b) commercial surgical mask (SM) taped (filled black symbols) and not taped (open black symbols) around manikin head at 30 L/min (upright triangles) and 85 L/min (inverted triangles).}
\label{fig9}
\end{figure}

\begin{figure*}
\begin{center}
\includegraphics[width=1.75\columnwidth]{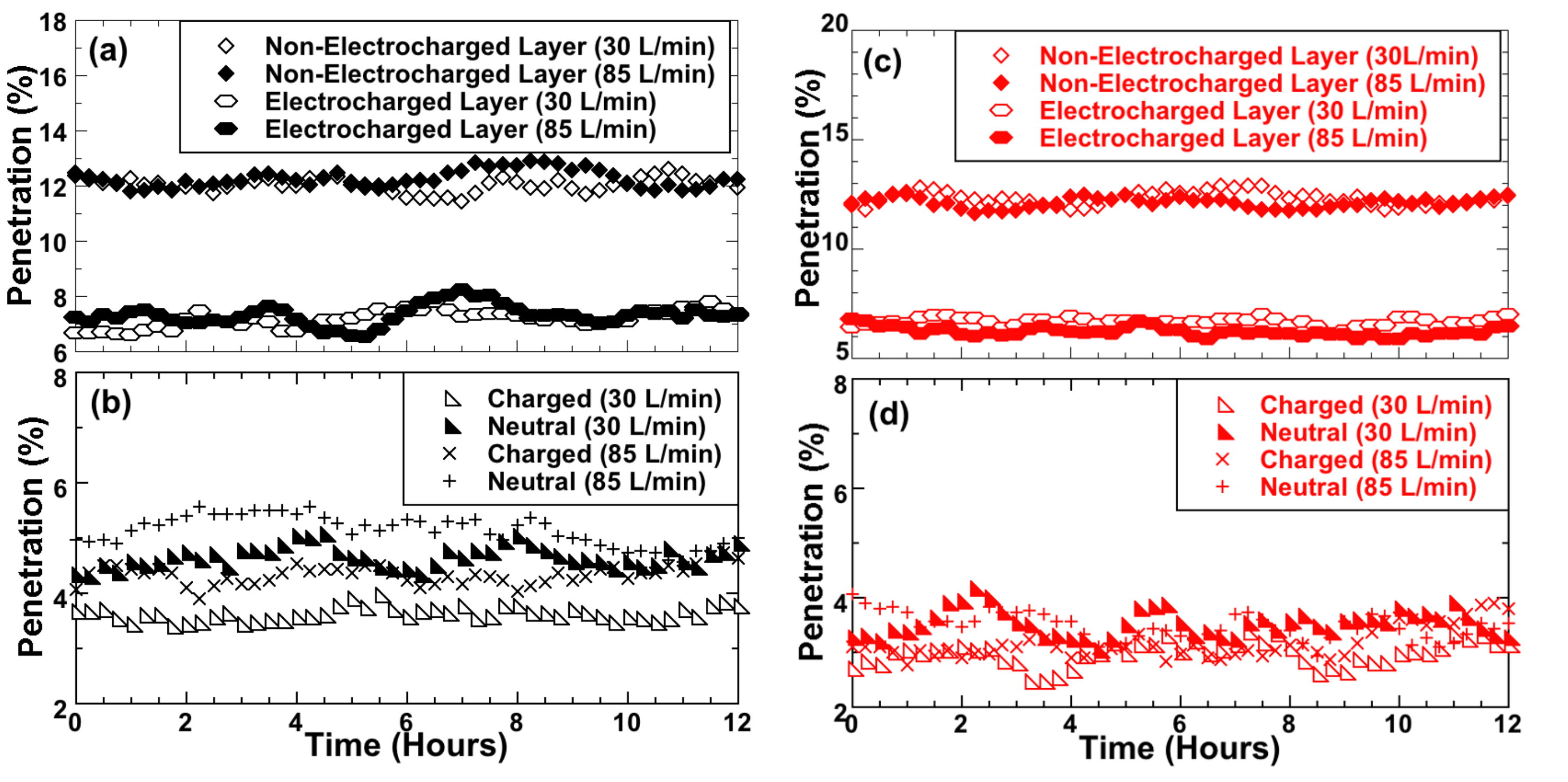}
\end{center}
\caption{Penetration P(t) (\%) versus time (hours) for (a) PP fabric mounted on a taped SM (b) PP fabric mounted on 3D printed Montana mask holder, (c) PP-PS fabric mounted on a taped SM, and (d) PP-PS fabric mounted on a 3D printed Montana mask holder.}
\label{fig10}
\end{figure*}

{\it Test Results:} FR filtration efficiency is usually measured in terms of the penetration percentage (P), defined as percentage of particles present in the environment that pass through the FRs and is quoted against the particle diameters. Since particle sizes could not be measured by the PM2.5 monitors, the penetration is instead defined as:
\begin{equation}
P(t) = \frac{C_B(t)}{C_A(t)} \times 100\%
\label{penetration}
\end{equation}
where $C_A(t)$ and $C_B(t)$ are the particle concentrations of PM2.5 monitors A and B respectively at time $t$. The penetration as a function of time was followed to study any deterioration in filtration properties. The PM2.5 monitors A and B were connected to a laptop and programmed to record concentration values at 15 minute intervals over a duration of 12 hours. Raw time series data for all filtration tests presented in fig.~\ref{fig9} and fig.~\ref{fig10} and Table~\ref{table3} are available at Ref.~\cite{BandiN95-Data1}.

The first tests were performed on commercial N95 FR and surgical masks (SMs) to obtain baseline calibration readings on the  filtration test setup designed in-house, which would then form the comparative standard for tests performed on filtration fabrics manufactured with the CC method. Since a tight facial fit for N95 FR is often emphasized, the test N95 FR was taped with a mask tape on to the manikin face and performed separate tests using aerosol particles that were both charged and charge neutral. On the other hand, baseline SM calibration tests were performed only with neutrally charged aerosol particles, but with and without mask tape applied for tight facial fit.  Figure~\ref{fig9} shows the results for the baseline commercial N95 FR and SM tests. A few details that are known among filtration research community are immediately apparent. Firstly, N95 FR filters out more than 95\% of the particles at both 30 and 85 L/min flow rates (see fig.~\ref{fig9}a). Secondly, a small but measurable difference in penetration percentage is clearly observable between charged (1 - 2\% penetration) and neutral ($\sim 3\%$ penetration) aerosol particles. Finally, no measurable dependence on flow rate could be observed for the N95 FR.

The SM results were in stark contrast with the N95 FR. Firstly, emphasis on the importance of tight facial fit becomes immediately apparent from fig.~\ref{fig9}b. When no tape was applied to close the interstitial gap between the face and mask, the penetration was nearly 50\% at 85 L/min and around 40\% at 35 L/min implying substantial leakage. Taping the mask around the manikin face drastically brought down the penetration percentage values to around 12\% at both flow rates, thereby underscoring the importance of tight facial fit in filtration masks and respirators. Secondly, even after applying tape to close the gap between face and mask, the penetration value of around 12\% was still higher than the N95 standard of $\le 5\%$. The reason for this discrepancy lies in the fact that SM fabrics do not employ electrocharged layers and rely entirely on inertial impaction and diffusion to achieve filtration, thus underscoring the importance of the electrocharged filtration mechanism.

Following these baseline tests, results are presented for tests performed on the two mask designs using isotactic PP fabrics. Figure~\ref{fig10}a presents results for the first design where a filtration fabric layer manufactured in-house was placed on the inner side of a commercial surgical mask (SM). Tests were  performed using an electrocharged layer as well as a fabric layer whose charge was depleted with a static eliminator as described earlier. All tests were performed by taping the SM to the manikin head. From fig.~\ref{fig10}a it is seen that commercial SM with non-electrocharged layer exhibits the same penetration percentage of around 12\% at both flow rates (30 and 85 L/min) and the results are statistically identical to ones shown in fig.~\ref{fig10}b for taped masks. However, adding the electrocharged filtration layer leads to a marked decrease in penetration of aerosol particles down to around 7\% at both flow rates. Unfortunately though, at 7\% penetration, it still falls short of the N95 FR requirement of $\le 5\%$. Be that as it may, at just 7\% penetration the proposed solution of an electrocharged fabric layer in a commercial SM provides ample protection for people using face masks in non-critical settings, but falls short of protection deemed necessary for say, emergency response and healthcare services.

Figure~\ref{fig10}b presents filtration test results for the 3D printed Montana mask design meant to achieve the N95 FR tight facial fit.  Four electrocharged filtration fabric layers of 0.2 - 0.3 mm thickness each were employed here. Charged aerosol particles exhibited slightly less penetration at 3.6\% (30 L/min) and 4.35\% (85 L/min) relative to neutral aerosols with 4.6\% (30 L/min) and 5.2\% (85 L/min). This trend is similar to one observed for commercial N95 FR in fig.~\ref{fig10}a, but the penetration values for fabrics manufactured using the CC method were unfortunately slightly higher than commerical N95 FR. The penetration values do fall within the 5\% requirement for the N95 standard for all cases except neutral aerosols at 85 L/min, which misses the target by a meagre 0.16\%. However, in natural work settings, one expects to be exposed to both charged and neutral aerosols. Therefore, when taken together, the mean penetration for all aerosols at 30 L/min flow rate is 4.12\% and at 85 L/min is 4.86\%, i.e. just below the 5\% penetration requirement. Ergo, the 3D printed Montana mask design fit with electrocharged filtration fabrics barely manages to meet the N95 standard, but it is emphasized once again that the test for pressure drop across the filtration fabrics could not be measured nor could the oscillatory inhalation-exhalation cycles present in normal respiration be simulated with the in-house filtration test setup.

Filtration tests were also performed under identical conditions using PP-PS fabrics. Figure~\ref{fig10}c shows results for SMs with and without PP-PS electrocharged fabric layer and should be compared against PP results in fig.~\ref{fig10}a. The penetration percentages were identical at both flow rates for the non-electrocharged layer, suggesting the slightly higher tortuosity in PP-PS fabric (fig.~\ref{fig7}c) relative to PP fabric (fig.~\ref{fig7}b) had no measurable impact on filtration quality. However, SM with PP-PS electrocharged filtration layer performed marginally better by $\sim 1\%$ (fig.~\ref{fig10}c compared to identical test with PP fabric (fig.~\ref{fig10}a). A similar improvement in penetration percentage was also observed for the 3D printed Montana mask design (see fig.~\ref{fig10}d). Tests for both charged and neutral aerosols at 30 as well as 85 L/min flow rate yielded penetration values in the same range of 3 - 4\%, modulo the variability one observes in the data (fig.~\ref{fig10}c). Ergo, the PP-PS fabrics surpassed the N95 requirement to within the testing limitations. Since the minor structural differences (tortuosity) between PP and PP-PS tests did not show any difference, it can only be surmised that the higher performance of PP-PS fabrics (relative to PP) comes from electrocharging as is confirmed in the following subsection. The mean penetration and standard deviation are listed for all our tests in Table~\ref{table3}, with all values rounded off to the second decimal place.

\begin{table*}
\caption{Penetration percentage values measured from filtration tests. Raw time series for all data are available at Ref.~\cite{BandiN95-Data1}}
\begin{tabular}{cc}
\hline
Material & Penetration (\%): Mean $\pm$ Stdev\\
\hline
N95 FR Charged (30 L/min) [Fig.~\ref{fig9}a] & 1.43 $\pm$ 0.03\\
N95 FR Neutral (30 L/min) [Fig.~\ref{fig9}a] & 3.29 $\pm$ 0.02\\
N95 FR Charged (85 L/min) [Fig.~\ref{fig9}a] & 1.54 $\pm$ 0.03\\
N95 FR Neutral (85 L/min [Fig.~\ref{fig9}a] &  3.37 $\pm$ 0.03\\
SM No Tape (30 L/min) [Fig.~\ref{fig9}b] & 40.72 $\pm$ 0.85\\
SM Taped (30 L/min) [Fig.~\ref{fig9}b] & 11.98 $\pm$ 0.73\\
SM No Tape (85 L/min) [Fig.~\ref{fig9}b] & 47.18 $\pm$ 0.45\\
SM Taped (85 L/min) [Fig.~\ref{fig9}b] & 13.13 $\pm$ 0.57\\
PP: SM + Non-Electrocharged Layer (30 L/min) [Fig.~\ref{fig10}a] & 12.04 $\pm$ 0.26\\
PP: SM + Non-Electrocharged Layer (85 L/min) [Fig.~\ref{fig10}a] & 12.26 $\pm$ 0.31\\
PP: SM + Electrocharged Layer (30 L/min) [Fig.~\ref{fig10}a] & 7.19 $\pm$ 0.32\\
PP: SM + Electrocharged Layer (85 L/min) [Fig.~\ref{fig10}a] & 7.3 $\pm$ 0.37\\
PP: 3D Charged (30 L/min) [Fig.~\ref{fig10}b] & 3.62 $\pm$ 0.13\\
PP: 3D Neutral (30 L/min) [Fig.~\ref{fig10}b] & 4.65 $\pm$ 0.2\\
PP: 3D Charged (85 L/min) [Fig.~\ref{fig10}b]  & 4.34 $\pm$ 0.17\\
PP: 3D Neutral (85 L/min) [Fig.~\ref{fig10}b] & 5.14 $\pm$ 0.27\\
PP-PS: SM + Non-Electrocharged Layer (30 L/min) [Fig.~\ref{fig10}c] & 12.31 $\pm$ 0.31\\
PP-PS: SM + Non-Electrocharged Layer (85 L/min) [Fig.~\ref{fig10}c] & 12.1 $\pm$ 0.24\\
PP-PS: SM + Electrocharged Layer (30 L/min) [Fig.~\ref{fig10}c] & 6.67 $\pm$ 0.19\\
PP-PS: SM + Electrocharged Layer (85 L/min) [Fig.~\ref{fig10}c] & 6.22 $\pm$ 0.22\\
PP-PS: 3D Charged (30 L/min) [Fig.~\ref{fig10}d] & 2.97 $\pm$ 0.22\\
PP-PS: 3D Neutral (30 L/min) [Fig.~\ref{fig10}d] & 3.53 $\pm$ 0.25\\
PP-PS: 3D Charged (85 L/min) [Fig.~\ref{fig10}d] & 3.17 $\pm$ 0.29\\
PP-PS: 3D Neutral (85 L/min) [Fig.~\ref{fig10}d] & 3.48 $\pm$ 0.25\\
\hline
\end{tabular}
\label{table3}
\end{table*}

{\it Nanoparticle filtration test: }An important question that arose was if the masks were capable of trapping individual viral particles, such as the SARS-Cov-2 virion. The SARS-Cov-2  virion is estimated to have a diameter between 50 - 200 nm. More precisely, Ref. \cite{Kim} places the SARS-Cov-2 virion of order 70 - 90 nm and Ref. \cite{Chen} places it at 50 - 200 nm. Although MPPS for N95 FRs is set at 300 nm (0.3 $\mu$m), some studies have suggested MPPS for N95 FRs occurs in the range of 40 - 60 nm \cite{Balazy, Rengasamy}. In particular, experiments by Balazy et al \cite{Balazy} using MS2 virion -- bacteriophage with single-stranded RNA comprising 3569 nucleotides that infects male {\it E. Coli} bacteria -- with an approximate diameter of 27.5 nm shows that N95 FRs can achieve superlative filtration for particle diameters smaller than its rated MPPS. This indicates that N95 FRs would perform equally well at trapping an individual SARS-Cov-2 virion with lower bound on diameter roughly 1.5 times that of MS2. 

Having placed the approximate range for the SARS-Cov-2 virion, the next question is if the SARS-Cov-2 virion possesses a surface electric charge that permits it to get attracted to an electrocharged filtration layer. The net charge would be determined by the sum of the charges exposed on the viron surface and can be calculated from the protein structure(s) at the surface. However, in order for the virus to be trapped by the electrocharged filters, there are two small issues to contend. Firstly, individual SARS-Cov-2 virions are not likely to be airborne. Instead, like most enveloped viruses, it is hydrated or in solution and is transmitted by aerosol droplets. When dehydrated, the virion's lipid membrane collapses and its proteins become denatured within a certain time, depending on temperature etc., i.e., it is rendered inactive.  Despite the low probability of encountering individual SARS-Cov-2 virions, it was a question still worthy of an investigative test.

Secondly, the net charge of a protein in solution (water, typically) depends on the pH. Every amino acid has a pKa (the log of the acid dissociation constant), which is an equilibrium constant indicating the pH at which the charges are balanced (net neutral). There are pKa values for each chemical group. If the pH is lower than the pKa, that amino acid becomes protonated, if it can. At low pH, there is an abundance of positively charged protons. Even some acids can become protonated at low pH. One can approximate the net charge of a linear peptide from its sequence and by providing a pH for the solvent using tools on the web (e.g. see Ref.~\cite{protcalc}) one can arrive at an estimate for surface charge. However, this does not take into account that structures are folded in 3D and some charges are hidden inside. Although it is only a rough value, it does still provide an estimate because residues buried inside are often hydrophobic, i.e. not charged.  There are several online tools available to calculate net charges of folded protein structures as well as their surface charge distribution, e.g. DELPHI \cite{delphi}. The isoelectric point pI is the pH at which the net charge is neutral. Below the pI, the net charge is positive, above it is negative. Ref.~\cite{Michen} lists some examples of viruses, most of their pIs being $< 7$, suggesting that most viruses would be net negatively charged at neutral pH, thus suggesting that the COVID-19 virion is also net negative. The main surface protein for SARS-Cov-2 is the spike protein \cite{spike}. It is glycosylated, thus has some amino acids, mostly arginines, chemically attached to sugars that can be charged, like sialic acid, which would also contribute to an overall negative charge. When the sequence of the spike 6VXX is input into Ref.~\cite{isoelectric}, a pI=5.8 is obtained, thus net negative.

Commercial filtration test systems permit detection of individual virion penetration as these systems can identify particle diameters. In the absence of such a system, and lacking knowledge of handling even inactive virions, a surrogate technique using fluorescent tagged nanoparticles was employed. Polystyrene nanospheres fluorescently tagged with Dragon Green (480 nm Absorption wavelength, 520 nm Emission wavelength) with mean diameter 50 nm $\pm$ 10 nm (Manufacturer Bangs Labs, Catalog No. FSDG001, Lot No. 14092) were used as a substitute for the SARS-Cov-2 virion. Since the polystyrene nanospheres were to mimic virions,  the aqueous solution was first dried and the nanoparticle dust was exposed to a brief burst of coronal discharge. Unsure whether coronal discharge exposure would dissociate fluorescent markers off the surface of the nanoparticles, a quick fluorescene test was performed and confirmed that the fluorescent nanoparticles performed to specification.

\begin{figure}
\begin{center}
\includegraphics[width=\columnwidth]{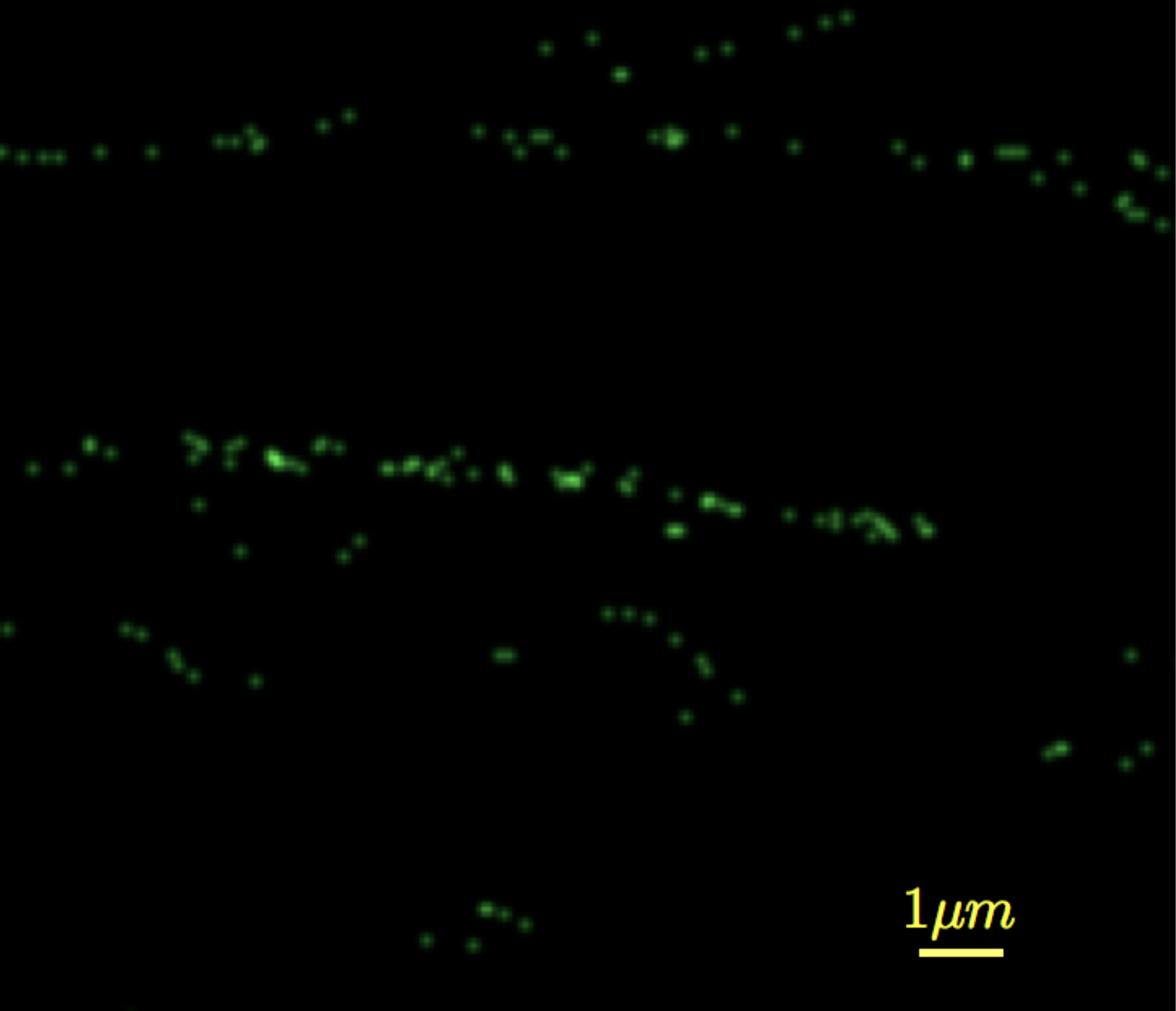}
\end{center}
\caption{(Color online) Confocal image of fluorescence emission signal from polystyrene nanospheres trapped along fiber surfaces at a depth of 500 nm from the outer surface of the electrocharged filtration layer exposed to the environment.}
\label{fig11}
\end{figure}

The filtration test was performed on the 3D mask design with nanoparticle dust replacing the mist generator in fig.~\ref{fig9}. Measurement of penetration was meaningless as the PM2.5 monitors could detect dust particles from ambient air entering the confining box. Instead, the test was conducted at 30 L/min flow rate for one hour duration and then the electrocharged filtration layers were interrogated under a confocal microscope (Nikon Andor Revolution WD spinning disk confocal, Laser wavelength 455 nm) for fluorescence signals from the trapped nanoparticles, if any. Figure~\ref{fig11} presents the fluorescence emission signal  from polystyrene particles trapped along fiber surfaces of the electrocharged layer as measured under the confocal at a depth of 500 nm from the outer surface (exposed to the environment). It is observed that the fluorescence signal occurs in clusters because the nanoparticles were not dispersed in aqueous solution as usually done in experiments. Instead the solution was dried to obtain nanodust that comprised clumps of particles, or alternatively it is possible the charged particles were preferentially attracted to charge centers that acted as traps and begs further future investigation. This was a proof of principle test to check whether the electrocharged fabrics were capable of trapping charged particles of dimensions comparable to the SARS-Cov-2 virion, and further quantitative analysis was not undertaken for the purposes of the current study.

\subsection{Charge}
Electrostatic charge measurements were made with a non-contact electrostatic potentiometer (KSS-2000 Digital Electrostatic Potentiometer, Manufacturer: Kasuga Denki Inc.) with a piezoelectric transducer capable of measuring electrostatic voltage at a distance of 50 -100 nm from the sample. The measurement readings for various fabric samples are listed in Table~\ref{table4}. The mean and standard deviation are taken over a set of 100 measurements conducted at various locations on several samples of each fabric material.

\begin{table*}
\caption{Electrostatic potential of various fabric samples.}
\begin{tabular}{cc}
\hline
Material &  Electrostatic potential (kV)\\
\hline
N95 FR & 11.1 $\pm$ 0.9\\
SM & 0.4 $\pm$ 0.3\\
PP & 5.3 $\pm$ 0.44\\
Isothermal charged PP & 7.4 $\pm$ 0.23\\
PS & 8.1 $\pm$ 0.28\\
Isothermal charged PS & 10.8 $\pm$ 0.41\\
PP-PS & 8.4 $\pm$ 0.36\\
Isothermal charged PP-PS & 10.1 $\pm$ 0.3\\
\hline
\end{tabular}
\label{table4}
\end{table*}

The commercial N95 FR fabric material which serves as the comparative standard for this study clearly holds the highest electrostatic potential at 11.1 $\pm$ 0.9 kV (see Table~\ref{table4}. The electrostatic potential quoted here for the commercial N95 FR was measured after extracting the electrocharged fabric layer sandwiched between the neutral outer layers. For this reason, the value quoted here will be higher than the actual potential an aerosol or dust particle experiences because of the outer neutral layer forming a dielectric barrier between the electrocharged layer and ambient environment. Be that as it may, for purposes of this study, the electrostatic potential of the commercial N95 FR's bare electrocharged layer forms the right comparative standard for fabrics manufactured with the CC method. SM fabrics had the lowest electrostatic potential at 0.4 $\pm$ 0.3 kV (see Table~\ref{table4}) and were practically neutral given the standard deviation (0.3 kV) nearly equals the mean (0.4 kV) electrostatic potential.

With the commercial N95 FR's electrocharged layer and SM fabrics as the two bounds of the comparative standard, it becomes possible to establish a correspondence between the electrostatic potential and filtration efficiency for all the fabrics studied here. Firstly, the measured electrostatic potential listed in Table~\ref{table4} shows PS fabrics possess more charge than isotactic PP fabrics, as is established by prior independent measurements \cite{PPPScharge}. Although both PP and PS are poor conductors in pure form, PS molecules have benzene rings which improve charge concentration of PS molecules \cite{PolyHBook}, this is not so for PP molecules which do not possess the benzene ring. However, several extraneous factors such as purity, thermal treatment (recrystallization in PP versus physical aging in PS) etc. also influence charge retention. Indeed, it is known that N95 FR electrocharged layers embed dopants \cite{dopePP1, dopePP2} for charge enhancement, particularly in polypropylene. Fabrication with simple materials and methods being the design strategy, such charge enhancement using dopants was not explored in the present work.

Secondly, isothermal charging of the manufactured fabric certainly increases the electrostatic potential of all samples by roughly 25\% (for PP-PS) to  40\% (for isotactic PP) thereby establishing the efficacy of the secondary isothermal charging step employed in the present study. Finally, a one-to-one correspondence is observed between electrostatic potential and filtration efficiency obtained from the penetration tests. The difference in filtration efficiency we observed between the PP and PP-PS samples could not be attributed to slight structural difference in tortuosity of the fiber porous matrix, but PP-PS fabrics hold higher electrostatic potential than PP fibers. This explains why PP-PS 3D printed Montana mask respirators consistently met the N95 requirement of $\le 5\%$ penetration, whereas PP 3D printed Montana masks could barely meet the requirement, with especially neutral particles falling slightly above the requirement.

\section{Conclusions}
In summary, a set of general design principles have been presented to construct one's own fabrication setup and manufacture of electrocharged filtration fabrics using the CC principle, together with some potential designs to use these filtration fabrics in face masks. In doing so, attention has been paid to utilising commonly available materials and easily replicable methods using two commonly available candidate materials, {\it viz.} PP and PS and their blends.

Electrocharged filtration fabrics must meet two requirements, structural heterogeneity for filtration through inertial impaction and diffusion, and high charge retention for electrocharged filtration. Fabric heterogeneity comes in several forms and includes a non-woven fabric comprised of fibers enmeshed in a random manner, fiber tortuosity which affects the fabric porosity, and variability of fiber radii which control the surface to volume ratio. The CC method can easily achieve a disordered non-woven fabric of enmeshed fibers, but tortuosity and variability of fiber radii is a material dependent parameter that enters {\it via} operating temperature. Whereas isotactic PP fabrics exhibited all three forms of heterogeneity (fiber radius 4.54 $\pm$ 6.2 $\mu$m), they possessed less charge (5.3 $\pm$ 0.44 kV without and 7.4 $\pm$ 0.23 kV with isothermal charging) relative to PS fabrics (8.1 $\pm$ 0.28 kV without and 10.8 $\pm$ 0.41 kV with isothermal charging). On the other hand PS fabrics resulted in thicker fiber radii of lower variability (18.5 $\pm$ 2.8 $\mu$m), albeit with higher tortuosity than PP fabrics. Achieving an optimal mix of heterogeneity and charge retention by blending PP and PS (fiber radii 4.9 $\pm$ 5.1 $\mu$m and charge of 8.4 $\pm$ 0.36 kV without and 10.1 $\pm$ 0.3 kV with isothermal charging) resulted in fabrics that were able to meet the N95 FR standard (fiber radii 4.1 $\pm$ 4.7 $\mu$m and electrostatic potential 11.1 $\pm$ 0.9 kV), at least within the limitations of the filtration test setup designed and constructed in-house.

The current work  maintained exclusive focus towards providing a proof of principle for the process and underlying mechanics of the CC principle as a viable method for fabrication of electrocharged filtration fabrics. Further work to be explored in future includes researching more commonly available fabric materials, such as Poly Ethylene Terepthalate, universally available as PET bottles for beverages, and Biaxially-oriented Poly Ethylene Terepthalate (BoPET), better known as Mylar and available as plastic bags and food wrapping. A more detailed characterisation of the structural and electrocharging properties of fabrics obtained via the CC method is also needed. Another avenue that begs exploration is the introduction of metallic nanoparticles at the fabrication stage to embed them into fabrics for charge enhancement, and possibly even decontamination, without endangering users as these fabrics are meant to be worn around mouth and nose for safe respiration.

Although the COVID-19 pandemic provided the impetus for the current effort, face masks may well become a mainstay of human social interactions going forward. On order of  two viruses jump across from animals to humans per year \cite{VJump} with most animals exhibiting viral richness showing propensity for close human contact \cite{V2}. If even a small fraction of those viruses result in asymptomatic viral shedding in human exhalation (breathing, coughing, or sneezing) \cite{V1}, face mask protection at the population level becomes a necessary means of protection. Decentralized local manufacture of face masks with high filtration efficiency from commonly available materials and simple designs could potentially alleviate global supply chain disruptions during such times, as witnessed during the COVID-19 pandemic \cite{manup}. It is hoped that this effort will help communities with face mask protection during such pandemics.

\acknowledgments
This work was supported by the Nonlinear and Non-equilibrium Physics Unit, OIST Graduate University. MMB first learned about the cotton candy method from a passing remark by Dr. L. Mahadevan in 2010. MMB  acknowledges advice from Dr. M. Wolf on COVID-19 surface charge characteristics, Dr. S. Ghosh on using PM2.5 air quality monitors for filtration testing, and help from Dr. K. Deasy with 3D printing, N. Ishizu with Scanning Electron Microscopy and Dr. H. B. Kang for characterisation, and OIST Imaging section with Confocal Microscopy. Dr. S. Velankar is gratefully acknowledged for critical reading of the manuscript and several helpful suggestions for its improvement.

\bibliography{all}

\begin{thebibliography}{52}
\expandafter\ifx\csname natexlab\endcsname\relax\def\natexlab#1{#1}\fi
\expandafter\ifx\csname bibnamefont\endcsname\relax
  \def\bibnamefont#1{#1}\fi
\expandafter\ifx\csname bibfnamefont\endcsname\relax
  \def\bibfnamefont#1{#1}\fi
\expandafter\ifx\csname citenamefont\endcsname\relax
  \def\citenamefont#1{#1}\fi
\expandafter\ifx\csname url\endcsname\relax
  \def\url#1{\texttt{#1}}\fi
\expandafter\ifx\csname urlprefix\endcsname\relax\def\urlprefix{URL }\fi
\providecommand{\bibinfo}[2]{#2}
\providecommand{\eprint}[2][]{\url{#2}}

\bibitem[{CDC()}]{CDCFMask}
\bibinfo{howpublished}{\url{https://www.cdc.gov/coronavirus/2019-ncov/prevent-getting-sick/diy-cloth-face-coverings.html}}.

\bibitem[{NIO()}]{NIOSH-N95}
\bibinfo{howpublished}{\url{https://www.cdc.gov/niosh/npptl/topics/respirators/disp_part/N95list1.html}}.

\bibitem[{3M()}]{3M}
\bibinfo{howpublished}{\url{https://multimedia.3m.com/mws/media/1791500O/comparison-ffp2-kn95}\\
  \url{-n95-filtering-facepiece-respirator-classes-tb.pdf}}.

\bibitem[{\citenamefont{Rogalski et~al.}(2017)\citenamefont{Rogalski,
  Bastiaansen, and Peijs}}]{RSpinReview}
\bibinfo{author}{\bibfnamefont{J.~J.} \bibnamefont{Rogalski}},
  \bibinfo{author}{\bibfnamefont{C.~W.~M.} \bibnamefont{Bastiaansen}},
  \bibnamefont{and} \bibinfo{author}{\bibfnamefont{T.}~\bibnamefont{Peijs}},
  \bibinfo{journal}{Nanocomposites} \textbf{\bibinfo{volume}{3}},
  \bibinfo{pages}{97} (\bibinfo{year}{2017}).

\bibitem[{\citenamefont{Mellado et~al.}(2011)\citenamefont{Mellado, McIlwee,
  Badrossamay, Goss, Mahadevan, and Parker}}]{RJet2}
\bibinfo{author}{\bibfnamefont{P.}~\bibnamefont{Mellado}},
  \bibinfo{author}{\bibfnamefont{H.~A.} \bibnamefont{McIlwee}},
  \bibinfo{author}{\bibfnamefont{M.~R.} \bibnamefont{Badrossamay}},
  \bibinfo{author}{\bibfnamefont{J.~A.} \bibnamefont{Goss}},
  \bibinfo{author}{\bibfnamefont{L.}~\bibnamefont{Mahadevan}},
  \bibnamefont{and} \bibinfo{author}{\bibfnamefont{K.~K.}
  \bibnamefont{Parker}}, \bibinfo{journal}{Appl. Phys. Lett.}
  \textbf{\bibinfo{volume}{99}}, \bibinfo{pages}{203107}
  (\bibinfo{year}{2011}).

\bibitem[{\citenamefont{Revoir and Bien}(1995)}]{Revoir1995}
\bibinfo{author}{\bibfnamefont{W.~H.} \bibnamefont{Revoir}} \bibnamefont{and}
  \bibinfo{author}{\bibfnamefont{C.~T.} \bibnamefont{Bien}},
  \emph{\bibinfo{title}{Respiratory Protection Handbook}}
  (\bibinfo{publisher}{Lewis Publisher, New York}, \bibinfo{year}{1995}).

\bibitem[{NAS(2006)}]{NAS2006}
\emph{\bibinfo{title}{Reusability of Facemasks During an Influenza Pandemic:
  Facing the Flu}} (\bibinfo{publisher}{Institute of Medicine, The National
  Academies Press, Washington, D. C.}, \bibinfo{year}{2006}).

\bibitem[{\citenamefont{Konda et~al.}(2020)\citenamefont{Konda, Prakash, Moss,
  Schmoldt, Grant, and Guha}}]{Konda}
\bibinfo{author}{\bibfnamefont{A.}~\bibnamefont{Konda}},
  \bibinfo{author}{\bibfnamefont{A.}~\bibnamefont{Prakash}},
  \bibinfo{author}{\bibfnamefont{G.~A.} \bibnamefont{Moss}},
  \bibinfo{author}{\bibfnamefont{M.}~\bibnamefont{Schmoldt}},
  \bibinfo{author}{\bibfnamefont{G.~D.} \bibnamefont{Grant}}, \bibnamefont{and}
  \bibinfo{author}{\bibfnamefont{S.}~\bibnamefont{Guha}}, \bibinfo{journal}{ACS
  Nano} \textbf{\bibinfo{volume}{14}}, \bibinfo{pages}{6339}
  (\bibinfo{year}{2020}).

\bibitem[{\citenamefont{Davies et~al.}(2013)\citenamefont{Davies, Thompson,
  Giri, Kafatos, Walter, and Bennett}}]{Davies}
\bibinfo{author}{\bibfnamefont{A.}~\bibnamefont{Davies}},
  \bibinfo{author}{\bibfnamefont{K.~A.} \bibnamefont{Thompson}},
  \bibinfo{author}{\bibfnamefont{K.}~\bibnamefont{Giri}},
  \bibinfo{author}{\bibfnamefont{G.}~\bibnamefont{Kafatos}},
  \bibinfo{author}{\bibfnamefont{J.}~\bibnamefont{Walter}}, \bibnamefont{and}
  \bibinfo{author}{\bibfnamefont{A.}~\bibnamefont{Bennett}},
  \bibinfo{journal}{Disaster Med. Public Health Prep.}
  \textbf{\bibinfo{volume}{7}}, \bibinfo{pages}{413} (\bibinfo{year}{2013}).

\bibitem[{\citenamefont{Frederick}(1974)}]{Frederick1974}
\bibinfo{author}{\bibfnamefont{E.~R.} \bibnamefont{Frederick}},
  \bibinfo{journal}{J. Air Pollution Control Assoc.}
  \textbf{\bibinfo{volume}{23}}, \bibinfo{pages}{1164} (\bibinfo{year}{1974}).

\bibitem[{\citenamefont{Teo and Ramakrishna}(2006)}]{ESpinRev1}
\bibinfo{author}{\bibfnamefont{W.~E.} \bibnamefont{Teo}} \bibnamefont{and}
  \bibinfo{author}{\bibfnamefont{S.}~\bibnamefont{Ramakrishna}},
  \bibinfo{journal}{Nanotechnology} \textbf{\bibinfo{volume}{17}},
  \bibinfo{pages}{R89} (\bibinfo{year}{2006}).

\bibitem[{\citenamefont{Pham et~al.}(2006)\citenamefont{Pham, Sharma, and
  Mikos}}]{ESpinRev2}
\bibinfo{author}{\bibfnamefont{Q.~P.} \bibnamefont{Pham}},
  \bibinfo{author}{\bibfnamefont{U.}~\bibnamefont{Sharma}}, \bibnamefont{and}
  \bibinfo{author}{\bibfnamefont{A.~G.} \bibnamefont{Mikos}},
  \bibinfo{journal}{Tissue Engg.} \textbf{\bibinfo{volume}{12}},
  \bibinfo{pages}{1197} (\bibinfo{year}{2006}).

\bibitem[{\citenamefont{Huang et~al.}(2003)\citenamefont{Huang, Zhang, Kotaki,
  and Ramakrishna}}]{ESpinRev3}
\bibinfo{author}{\bibfnamefont{Z.~M.} \bibnamefont{Huang}},
  \bibinfo{author}{\bibfnamefont{Y.~Z.} \bibnamefont{Zhang}},
  \bibinfo{author}{\bibfnamefont{M.}~\bibnamefont{Kotaki}}, \bibnamefont{and}
  \bibinfo{author}{\bibfnamefont{S.}~\bibnamefont{Ramakrishna}},
  \bibinfo{journal}{Composites Sci. and Tech.} \textbf{\bibinfo{volume}{63}},
  \bibinfo{pages}{2223} (\bibinfo{year}{2003}).

\bibitem[{\citenamefont{Schiffman and Schauer}(2008)}]{ESpinRev4}
\bibinfo{author}{\bibfnamefont{J.~D.} \bibnamefont{Schiffman}}
  \bibnamefont{and} \bibinfo{author}{\bibfnamefont{C.~L.}
  \bibnamefont{Schauer}}, \bibinfo{journal}{Polymer Reviews}
  \textbf{\bibinfo{volume}{48}}, \bibinfo{pages}{317} (\bibinfo{year}{2008}).

\bibitem[{\citenamefont{Chronakis}(2005)}]{ESpinRev5}
\bibinfo{author}{\bibfnamefont{I.~S.} \bibnamefont{Chronakis}},
  \bibinfo{journal}{J. Mat. Proc. Tech.} \textbf{\bibinfo{volume}{167}},
  \bibinfo{pages}{283} (\bibinfo{year}{2005}).

\bibitem[{\citenamefont{Hutmacher and Dalton}(2011)}]{MESpin1}
\bibinfo{author}{\bibfnamefont{D.~M.} \bibnamefont{Hutmacher}}
  \bibnamefont{and} \bibinfo{author}{\bibfnamefont{P.~D.}
  \bibnamefont{Dalton}}, \bibinfo{journal}{Chemistry}
  \textbf{\bibinfo{volume}{6}}, \bibinfo{pages}{44} (\bibinfo{year}{2011}).

\bibitem[{\citenamefont{Dalton et~al.}(2007)\citenamefont{Dalton, Grafahrend,
  Klinkhammer, Klee, and M{\"o}ller}}]{MESpin2}
\bibinfo{author}{\bibfnamefont{P.~D.} \bibnamefont{Dalton}},
  \bibinfo{author}{\bibfnamefont{D.}~\bibnamefont{Grafahrend}},
  \bibinfo{author}{\bibfnamefont{K.}~\bibnamefont{Klinkhammer}},
  \bibinfo{author}{\bibfnamefont{D.}~\bibnamefont{Klee}}, \bibnamefont{and}
  \bibinfo{author}{\bibfnamefont{M.}~\bibnamefont{M{\"o}ller}},
  \bibinfo{journal}{Polymer} \textbf{\bibinfo{volume}{48}},
  \bibinfo{pages}{6823} (\bibinfo{year}{2007}).

\bibitem[{\citenamefont{Pu et~al.}(2018)\citenamefont{Pu, Zheng, Chen, Long,
  Wu, Li, Wang, and Ning}}]{MBlowing1}
\bibinfo{author}{\bibfnamefont{Y.}~\bibnamefont{Pu}},
  \bibinfo{author}{\bibfnamefont{J.}~\bibnamefont{Zheng}},
  \bibinfo{author}{\bibfnamefont{F.}~\bibnamefont{Chen}},
  \bibinfo{author}{\bibfnamefont{Y.}~\bibnamefont{Long}},
  \bibinfo{author}{\bibfnamefont{H.}~\bibnamefont{Wu}},
  \bibinfo{author}{\bibfnamefont{Q.}~\bibnamefont{Li}},
  \bibinfo{author}{\bibfnamefont{X.}~\bibnamefont{Wang}}, \bibnamefont{and}
  \bibinfo{author}{\bibfnamefont{X.}~\bibnamefont{Ning}},
  \bibinfo{journal}{Polymers (Basel)} \textbf{\bibinfo{volume}{10}},
  \bibinfo{pages}{959} (\bibinfo{year}{2018}).

\bibitem[{\citenamefont{Zhang et~al.}(2018{\natexlab{a}})\citenamefont{Zhang,
  Liu, Zhang, Huang, and Jin}}]{MBlowing2}
\bibinfo{author}{\bibfnamefont{H.}~\bibnamefont{Zhang}},
  \bibinfo{author}{\bibfnamefont{J.}~\bibnamefont{Liu}},
  \bibinfo{author}{\bibfnamefont{X.}~\bibnamefont{Zhang}},
  \bibinfo{author}{\bibfnamefont{C.}~\bibnamefont{Huang}}, \bibnamefont{and}
  \bibinfo{author}{\bibfnamefont{X.}~\bibnamefont{Jin}}, \bibinfo{journal}{RSC
  Advances} \textbf{\bibinfo{volume}{198}}, \bibinfo{pages}{7932}
  (\bibinfo{year}{2018}{\natexlab{a}}).

\bibitem[{\citenamefont{Drabek and Zatloukal}(2019)}]{MBlowing3}
\bibinfo{author}{\bibfnamefont{J.}~\bibnamefont{Drabek}} \bibnamefont{and}
  \bibinfo{author}{\bibfnamefont{M.}~\bibnamefont{Zatloukal}},
  \bibinfo{journal}{Phys. Fluids} \textbf{\bibinfo{volume}{31}},
  \bibinfo{pages}{091301} (\bibinfo{year}{2019}).

\bibitem[{\citenamefont{Kellie}(2016)}]{MBlowBook1}
\bibinfo{author}{\bibfnamefont{G.}~\bibnamefont{Kellie}},
  \emph{\bibinfo{title}{Advances in Technical Nonwovens}}
  (\bibinfo{publisher}{Woodhead Publishing, Elsevier Ltd.},
  \bibinfo{year}{2016}).

\bibitem[{\citenamefont{Zhang}(2014)}]{MBlowBook2}
\bibinfo{author}{\bibfnamefont{D.}~\bibnamefont{Zhang}},
  \emph{\bibinfo{title}{Advances in Filament Yarn Spinning of Textiles and
  Polymers}} (\bibinfo{publisher}{Cambridge, UK: Woodhead Publishing Ltd.},
  \bibinfo{year}{2014}).

\bibitem[{\citenamefont{Badrossamay et~al.}(2010)\citenamefont{Badrossamay,
  McIlwee, Goss, and Parker}}]{RJet1}
\bibinfo{author}{\bibfnamefont{M.~R.} \bibnamefont{Badrossamay}},
  \bibinfo{author}{\bibfnamefont{H.~A.} \bibnamefont{McIlwee}},
  \bibinfo{author}{\bibfnamefont{J.~A.} \bibnamefont{Goss}}, \bibnamefont{and}
  \bibinfo{author}{\bibfnamefont{K.~K.} \bibnamefont{Parker}},
  \bibinfo{journal}{Nano Lett.} \textbf{\bibinfo{volume}{10}},
  \bibinfo{pages}{2257} (\bibinfo{year}{2010}).

\bibitem[{\citenamefont{Rogalski et~al.}(2018)\citenamefont{Rogalski,
  Bastiaansen, and Peijs}}]{ES-RSComp}
\bibinfo{author}{\bibfnamefont{J.~J.} \bibnamefont{Rogalski}},
  \bibinfo{author}{\bibfnamefont{C.~W.~M.} \bibnamefont{Bastiaansen}},
  \bibnamefont{and} \bibinfo{author}{\bibfnamefont{T.}~\bibnamefont{Peijs}},
  \bibinfo{journal}{Fibers} \textbf{\bibinfo{volume}{6}}, \bibinfo{pages}{37}
  (\bibinfo{year}{2018}).

\bibitem[{\citenamefont{Wang et~al.}(2011)\citenamefont{Wang, Shi, Secret, and
  Chen}}]{CCandy1}
\bibinfo{author}{\bibfnamefont{L.}~\bibnamefont{Wang}},
  \bibinfo{author}{\bibfnamefont{J.}~\bibnamefont{Shi}},
  \bibinfo{author}{\bibfnamefont{E.}~\bibnamefont{Secret}}, \bibnamefont{and}
  \bibinfo{author}{\bibfnamefont{Y.}~\bibnamefont{Chen}},
  \bibinfo{journal}{Microelectronic Engg.} \textbf{\bibinfo{volume}{88}},
  \bibinfo{pages}{1718} (\bibinfo{year}{2011}).

\bibitem[{\citenamefont{Wongpajan et~al.}(2018)\citenamefont{Wongpajan,
  Thumsorn, Inoya, Okoshi, and Hamada}}]{CCandy2}
\bibinfo{author}{\bibfnamefont{R.}~\bibnamefont{Wongpajan}},
  \bibinfo{author}{\bibfnamefont{S.}~\bibnamefont{Thumsorn}},
  \bibinfo{author}{\bibfnamefont{H.}~\bibnamefont{Inoya}},
  \bibinfo{author}{\bibfnamefont{M.}~\bibnamefont{Okoshi}}, \bibnamefont{and}
  \bibinfo{author}{\bibfnamefont{H.}~\bibnamefont{Hamada}},
  \bibinfo{journal}{Fibers and Polymers} \textbf{\bibinfo{volume}{19}},
  \bibinfo{pages}{135} (\bibinfo{year}{2018}).

\bibitem[{\citenamefont{Choi et~al.}(1988)\citenamefont{Choi, Spruiell,
  Fellers, and Wadsworth}}]{Strength}
\bibinfo{author}{\bibfnamefont{K.~J.} \bibnamefont{Choi}},
  \bibinfo{author}{\bibfnamefont{J.~E.} \bibnamefont{Spruiell}},
  \bibinfo{author}{\bibfnamefont{J.~F.} \bibnamefont{Fellers}},
  \bibnamefont{and} \bibinfo{author}{\bibfnamefont{L.~C.}
  \bibnamefont{Wadsworth}}, \bibinfo{journal}{Polym. Eng. Sci}
  \textbf{\bibinfo{volume}{28}}, \bibinfo{pages}{81} (\bibinfo{year}{1988}).

\bibitem[{\citenamefont{Lee and Wadsworth}(1990)}]{StrucFilt}
\bibinfo{author}{\bibfnamefont{Y.}~\bibnamefont{Lee}} \bibnamefont{and}
  \bibinfo{author}{\bibfnamefont{L.~C.} \bibnamefont{Wadsworth}},
  \bibinfo{journal}{Polym. Eng. Sci.} \textbf{\bibinfo{volume}{30}},
  \bibinfo{pages}{1413} (\bibinfo{year}{1990}).

\bibitem[{\citenamefont{Sessler}(1998)}]{Electrets}
\bibinfo{author}{\bibfnamefont{G.~M.} \bibnamefont{Sessler}},
  \emph{\bibinfo{title}{Electrets. Vol. 1}} (\bibinfo{publisher}{Laplacian
  ress: Morgan Hill, California}, \bibinfo{year}{1998}).

\bibitem[{\citenamefont{Giacometti and Oliveira}(1992)}]{coronachg1}
\bibinfo{author}{\bibfnamefont{J.~A.} \bibnamefont{Giacometti}}
  \bibnamefont{and} \bibinfo{author}{\bibfnamefont{O.~N.}
  \bibnamefont{Oliveira}}, \bibinfo{journal}{IEEE Trans. Elec. Insulation}
  \textbf{\bibinfo{volume}{27}}, \bibinfo{pages}{924} (\bibinfo{year}{1992}).

\bibitem[{\citenamefont{Kumara et~al.}(2009)\citenamefont{Kumara, Serdyuk, and
  Gubanski}}]{coronachg2}
\bibinfo{author}{\bibfnamefont{S.}~\bibnamefont{Kumara}},
  \bibinfo{author}{\bibfnamefont{Y.~V.} \bibnamefont{Serdyuk}},
  \bibnamefont{and} \bibinfo{author}{\bibfnamefont{S.~M.}
  \bibnamefont{Gubanski}}, \bibinfo{journal}{IEEE Trans. Dielectrics and Elec.
  Insulation} \textbf{\bibinfo{volume}{16}}, \bibinfo{pages}{726}
  (\bibinfo{year}{2009}).

\bibitem[{\citenamefont{Molini{\'e}}(1999)}]{CoronaPP1}
\bibinfo{author}{\bibfnamefont{P.}~\bibnamefont{Molini{\'e}}},
  \bibinfo{journal}{J. Electrostat.} \textbf{\bibinfo{volume}{45}},
  \bibinfo{pages}{265} (\bibinfo{year}{1999}).

\bibitem[{\citenamefont{Kravtsov et~al.}(2000)\citenamefont{Kravtsov,
  Br{\"u}nig, Zhandarov, and Beyreuther}}]{CoronaPP2}
\bibinfo{author}{\bibfnamefont{A.}~\bibnamefont{Kravtsov}},
  \bibinfo{author}{\bibfnamefont{H.}~\bibnamefont{Br{\"u}nig}},
  \bibinfo{author}{\bibfnamefont{S.}~\bibnamefont{Zhandarov}},
  \bibnamefont{and}
  \bibinfo{author}{\bibfnamefont{R.}~\bibnamefont{Beyreuther}},
  \bibinfo{journal}{Adv. Polymer Tech.} \textbf{\bibinfo{volume}{19}},
  \bibinfo{pages}{312} (\bibinfo{year}{2000}).

\bibitem[{MMa()}]{MMask}
\bibinfo{howpublished}{\url{https://www.makethemasks.com/3d-printing}}.

\bibitem[{\citenamefont{Bandi}(2020)}]{BandiN95-Data1}
\bibinfo{author}{\bibfnamefont{M.~M.} \bibnamefont{Bandi}},
  \bibinfo{journal}{Dryad, Data set
  \url{https://doi.org/10.5061/dryad.ffbg79crq}}  (\bibinfo{year}{2020}).

\bibitem[{\citenamefont{Kim et~al.}(2020)\citenamefont{Kim, Chung, Jo, Lee,
  Kim, Woo, Park, Kim, Kim, and Han}}]{Kim}
\bibinfo{author}{\bibfnamefont{J.-M.} \bibnamefont{Kim}},
  \bibinfo{author}{\bibfnamefont{Y.-S.} \bibnamefont{Chung}},
  \bibinfo{author}{\bibfnamefont{H.~J.} \bibnamefont{Jo}},
  \bibinfo{author}{\bibfnamefont{N.-J.} \bibnamefont{Lee}},
  \bibinfo{author}{\bibfnamefont{M.~S.} \bibnamefont{Kim}},
  \bibinfo{author}{\bibfnamefont{S.~H.} \bibnamefont{Woo}},
  \bibinfo{author}{\bibfnamefont{S.}~\bibnamefont{Park}},
  \bibinfo{author}{\bibfnamefont{J.~W.} \bibnamefont{Kim}},
  \bibinfo{author}{\bibfnamefont{H.~M.} \bibnamefont{Kim}}, \bibnamefont{and}
  \bibinfo{author}{\bibfnamefont{M.-G.} \bibnamefont{Han}},
  \bibinfo{journal}{Oosong Public Health and Res. Perspect.}
  \textbf{\bibinfo{volume}{11}}, \bibinfo{pages}{3} (\bibinfo{year}{2020}).

\bibitem[{\citenamefont{Chen et~al.}(2020)\citenamefont{Chen, Zhou, Dong, Qu,
  Gong, Han, Qiu, Wang, Liu, Wei et~al.}}]{Chen}
\bibinfo{author}{\bibfnamefont{N.}~\bibnamefont{Chen}},
  \bibinfo{author}{\bibfnamefont{M.}~\bibnamefont{Zhou}},
  \bibinfo{author}{\bibfnamefont{X.}~\bibnamefont{Dong}},
  \bibinfo{author}{\bibfnamefont{J.}~\bibnamefont{Qu}},
  \bibinfo{author}{\bibfnamefont{F.}~\bibnamefont{Gong}},
  \bibinfo{author}{\bibfnamefont{Y.}~\bibnamefont{Han}},
  \bibinfo{author}{\bibfnamefont{Y.}~\bibnamefont{Qiu}},
  \bibinfo{author}{\bibfnamefont{J.}~\bibnamefont{Wang}},
  \bibinfo{author}{\bibfnamefont{Y.}~\bibnamefont{Liu}},
  \bibinfo{author}{\bibfnamefont{Y.}~\bibnamefont{Wei}}, \bibnamefont{et~al.},
  \bibinfo{journal}{Lancet} \textbf{\bibinfo{volume}{395}},
  \bibinfo{pages}{507} (\bibinfo{year}{2020}).

\bibitem[{\citenamefont{Balazy et~al.}(2006)\citenamefont{Balazy, Toivola,
  Reponen, Podg{\'o}rski, Zimmer, and Grinshpun}}]{Balazy}
\bibinfo{author}{\bibfnamefont{A.}~\bibnamefont{Balazy}},
  \bibinfo{author}{\bibfnamefont{M.}~\bibnamefont{Toivola}},
  \bibinfo{author}{\bibfnamefont{T.}~\bibnamefont{Reponen}},
  \bibinfo{author}{\bibfnamefont{A.}~\bibnamefont{Podg{\'o}rski}},
  \bibinfo{author}{\bibfnamefont{A.}~\bibnamefont{Zimmer}}, \bibnamefont{and}
  \bibinfo{author}{\bibfnamefont{S.~A.} \bibnamefont{Grinshpun}},
  \bibinfo{journal}{Ann. of Occup. Hygiene} \textbf{\bibinfo{volume}{50}},
  \bibinfo{pages}{259} (\bibinfo{year}{2006}).

\bibitem[{\citenamefont{Rengasamy et~al.}(2007)\citenamefont{Rengasamy,
  Verbofsky, King, and Shaffer}}]{Rengasamy}
\bibinfo{author}{\bibfnamefont{S.}~\bibnamefont{Rengasamy}},
  \bibinfo{author}{\bibfnamefont{R.}~\bibnamefont{Verbofsky}},
  \bibinfo{author}{\bibfnamefont{W.~P.} \bibnamefont{King}}, \bibnamefont{and}
  \bibinfo{author}{\bibfnamefont{R.~E.} \bibnamefont{Shaffer}},
  \bibinfo{journal}{J. Int. Soc. Resp. Protection}
  \textbf{\bibinfo{volume}{24}}, \bibinfo{pages}{49} (\bibinfo{year}{2007}).

\bibitem[{pro()}]{protcalc}
\bibinfo{howpublished}{\url{http://protcalc.sourceforge.net/}}.

\bibitem[{del()}]{delphi}
\bibinfo{howpublished}{\url{http://honig.c2b2.columbia.edu/delphi}}.

\bibitem[{\citenamefont{Michen and Graule}(2010)}]{Michen}
\bibinfo{author}{\bibfnamefont{B.}~\bibnamefont{Michen}} \bibnamefont{and}
  \bibinfo{author}{\bibfnamefont{T.}~\bibnamefont{Graule}},
  \bibinfo{journal}{J. Appl. Microbiology} \textbf{\bibinfo{volume}{109}},
  \bibinfo{pages}{388} (\bibinfo{year}{2010}).

\bibitem[{spi()}]{spike}
\bibinfo{howpublished}{\url{http://www.rcsb.org/structure/6VXX}}.

\bibitem[{iso()}]{isoelectric}
\bibinfo{howpublished}{\url{ http://isoelectric.org/calculate.php}}.

\bibitem[{\citenamefont{Dodbiba et~al.}(2003)\citenamefont{Dodbiba, Shibayama,
  Miyazaki, and Fujita}}]{PPPScharge}
\bibinfo{author}{\bibfnamefont{G.}~\bibnamefont{Dodbiba}},
  \bibinfo{author}{\bibfnamefont{A.}~\bibnamefont{Shibayama}},
  \bibinfo{author}{\bibfnamefont{T.}~\bibnamefont{Miyazaki}}, \bibnamefont{and}
  \bibinfo{author}{\bibfnamefont{T.}~\bibnamefont{Fujita}},
  \bibinfo{journal}{Mat. Trans.} \textbf{\bibinfo{volume}{44}},
  \bibinfo{pages}{161} (\bibinfo{year}{2003}).

\bibitem[{\citenamefont{Wypych}(2016)}]{PolyHBook}
\bibinfo{author}{\bibfnamefont{G.}~\bibnamefont{Wypych}},
  \emph{\bibinfo{title}{Handbook of Polymers}} (\bibinfo{publisher}{Toronto,
  Canada: ChemTec Publishing; 2nd Edition}, \bibinfo{year}{2016}).

\bibitem[{\citenamefont{Zhang et~al.}(2018{\natexlab{b}})\citenamefont{Zhang,
  Liu, Zhang, Huang, and Jin}}]{dopePP1}
\bibinfo{author}{\bibfnamefont{H.}~\bibnamefont{Zhang}},
  \bibinfo{author}{\bibfnamefont{J.}~\bibnamefont{Liu}},
  \bibinfo{author}{\bibfnamefont{X.}~\bibnamefont{Zhang}},
  \bibinfo{author}{\bibfnamefont{C.}~\bibnamefont{Huang}}, \bibnamefont{and}
  \bibinfo{author}{\bibfnamefont{X.}~\bibnamefont{Jin}}, \bibinfo{journal}{RSC
  Advances} \textbf{\bibinfo{volume}{8}}, \bibinfo{pages}{7932}
  (\bibinfo{year}{2018}{\natexlab{b}}).

\bibitem[{\citenamefont{Kilic et~al.}(2015)\citenamefont{Kilic, Shim, and
  Pourdeyhimi}}]{dopePP2}
\bibinfo{author}{\bibfnamefont{A.}~\bibnamefont{Kilic}},
  \bibinfo{author}{\bibfnamefont{E.}~\bibnamefont{Shim}}, \bibnamefont{and}
  \bibinfo{author}{\bibfnamefont{B.}~\bibnamefont{Pourdeyhimi}},
  \bibinfo{journal}{Aerosol Sci. and Tech.} \textbf{\bibinfo{volume}{49}},
  \bibinfo{pages}{666} (\bibinfo{year}{2015}).

\bibitem[{\citenamefont{Woolhouse et~al.}(2008)\citenamefont{Woolhouse, Howey,
  Gaunt, Reilly, Chase-Topping, and Savill}}]{VJump}
\bibinfo{author}{\bibfnamefont{M.~E.~J.} \bibnamefont{Woolhouse}},
  \bibinfo{author}{\bibfnamefont{R.}~\bibnamefont{Howey}},
  \bibinfo{author}{\bibfnamefont{E.}~\bibnamefont{Gaunt}},
  \bibinfo{author}{\bibfnamefont{L.}~\bibnamefont{Reilly}},
  \bibinfo{author}{\bibfnamefont{M.}~\bibnamefont{Chase-Topping}},
  \bibnamefont{and} \bibinfo{author}{\bibfnamefont{N.}~\bibnamefont{Savill}},
  \bibinfo{journal}{Proc. R. Soc. B.} \textbf{\bibinfo{volume}{275}},
  \bibinfo{pages}{2111} (\bibinfo{year}{2008}).

\bibitem[{\citenamefont{Olival et~al.}(2017)\citenamefont{Olival, Hosseini,
  Zambrana-Torrelio, Ross, Bogich, and Daszak}}]{V2}
\bibinfo{author}{\bibfnamefont{K.~J.} \bibnamefont{Olival}},
  \bibinfo{author}{\bibfnamefont{P.~R.} \bibnamefont{Hosseini}},
  \bibinfo{author}{\bibfnamefont{C.}~\bibnamefont{Zambrana-Torrelio}},
  \bibinfo{author}{\bibfnamefont{N.}~\bibnamefont{Ross}},
  \bibinfo{author}{\bibfnamefont{T.~L.} \bibnamefont{Bogich}},
  \bibnamefont{and} \bibinfo{author}{\bibfnamefont{P.}~\bibnamefont{Daszak}},
  \bibinfo{journal}{Nature} \textbf{\bibinfo{volume}{546}},
  \bibinfo{pages}{646} (\bibinfo{year}{2017}).

\bibitem[{\citenamefont{Fraser et~al.}(2004)\citenamefont{Fraser, Riley,
  Anderson, and Ferguson}}]{V1}
\bibinfo{author}{\bibfnamefont{C.}~\bibnamefont{Fraser}},
  \bibinfo{author}{\bibfnamefont{S.}~\bibnamefont{Riley}},
  \bibinfo{author}{\bibfnamefont{R.~M.} \bibnamefont{Anderson}},
  \bibnamefont{and} \bibinfo{author}{\bibfnamefont{N.~M.}
  \bibnamefont{Ferguson}}, \bibinfo{journal}{Proc. Nat. Acad. Sci. of the USA}
  \textbf{\bibinfo{volume}{101}}, \bibinfo{pages}{6146} (\bibinfo{year}{2004}).

\bibitem[{\citenamefont{Molina}(2020)}]{manup}
\bibinfo{author}{\bibfnamefont{A.~{\it et~al}.} \bibnamefont{Molina}},
  \bibinfo{journal}{arXiv:2004.13494 [physics.app-ph]}  (\bibinfo{year}{2020}).

\end{thebibliography}
\end{document}